 \documentclass[preprint,3p,12pt]{elsarticle}
\usepackage{amssymb}
\usepackage{graphicx,color,slashed,amsfonts,amsmath,mathrsfs}
\usepackage{hyperref}
\usepackage{subfig}
\usepackage{listings}
\usepackage[utf8]{inputenc}

\usepackage{amsthm}

\newcounter{bla}

\newcommand{\mgamc}{{\sc Mad\-Graph5\-\_aMC\-@\-NLO}}
\newcommand{\mg}{{\sc Mad\-Graph5\-\_aMC\-@\-NLO}}

\newcommand{\syscalc}{{\sc SysCalc}}
\newcommand{\pythia}{{\sc Pythia8}}
\newcommand{\pt}{{$P_{\rm T}$}}

\newcommand{\delphes}{{\sc Delphes}}

\begin{document}

\begin{frontmatter}

\title{The \syscalc\ code: A tool to derive theoretical systematic uncertainties }


\author[a]{Alexis Kalogeropoulos}
\author[b]{Johan Alwall}

\cortext[author] {Corresponding author.\\\textit{E-mail address:} Alexis.Kalogeropoulos@cern.ch}
\address[a]{Princeton University, Department of Physics, \\Jadwin Hall, Washington Road, Princeton, NJ 08544-0708} 
\address[b]{National Taiwan University,\\   No.1, Sec. 4, Roosevelt Road Taipei, 106, Taiwan (at the time of the development of the tool)}

\begin{abstract}

Undisputedly, derivation of theoretical systematic uncertainties is an
inseparable ingredient of any robust analysis dealing with
experimental data. However, it is not uncommon, even for those analyses
that use state of the art methods and tools to suffer from insufficient
statistics when it comes to the simulated datasets used to estimate systematic
uncertainties. This practically limits the power, and sometimes the
robustness of the analysis. 
\\
 
In this paper, we present \syscalc, a code which is able to derive
weights for various important theoretical systematic uncertainties,
including those related to the choice of the Parton Distribution Function sets and the various
scale choices. \syscalc\ utilizes the central sample generated events to estimate the related systematic uncertainties, 
thus, omitting the need for generating dedicated systematics datasets, and with only a minimal added cost in terms of computing resources. 
In this paper we discuss the working principles of the
code accompanied by various validation plots. We also discuss the structure
of the code followed by a practical guide for how to use the tool.
\\

\end{abstract}


\end{frontmatter}

% \linenumbers

\newpage

\section{Introduction}

In the high-energy physics (HEP) community, it is a common practise to choose a scale  for the normalization ($\mu_R$), the factorization scale  ($\mu_F$), the emission scale ($\alpha_s$) as well as a choice for the Parton Distribution Function (PDF), and then generate large simulated datasets serving as the central samples. This is the case for example for both the ATLAS~\cite{atlas} and the CMS~\cite{cms} Collaborations, which typically generate billions of Monte Carlo (MC) simulated events so as to make sure that a sufficient estimation of the physical processes under question with the usage of the scales and PDFs that presumably best describe data, can be attained. The above come with a non negligible cost in terms of resources (both human and computing) but also on time spent to not only design, test and validate, but also to complete the generation of such simulated samples which have to be big enough in order to achieve small statistical uncertainties at the same time.  %Of course, in any Monte Carlo (MC) large scale generation, the preferred choice is to use a set of options which could pottentially describe data better. 
\\

Moreover, one could argue that more than one set of parameters (which is not always trivial to identify) can exist which should be modelled  coherently as well. It is clear that the HEP community (both experimentalists and theoreticians) needs to have access to simulated events generated with different settings, scales, or both. Nowadays, it is common practice that analyses have to estimate systematic uncertainties as well, where some of the most important theoretical ones include the variation of the aforementioned $\mu_R$, $\mu_F$, $\alpha_s$ or the usage of different PDF grid sets. But up to now, it was often very unpractical, or even impossible in many case, to generate datasets big enough in order to have comparable statistics to the central one(s). This results in high statistical uncertainties on the systematic uncertainties which constrain the power and accuracy of several important analyses, like for example the ones dealing with cross section or precision mass measurements for instance. Such examples can be found in~\cite{cms_an2,cms_an1} where the dominant uncertainty is the systematic one which can be directly related to limited statistics of the associated simulated systematic datasets. 
\\

In this paper, we present a new tool, \syscalc, which is capable of providing to the end user various weights based on the nominal generated sample of a given physics process, in order to avoid the re-generation of events with different choices on the scales, the PDF grid sets etc. The main functionality of \syscalc\ is the computation of weights for various systematic variations (scales, PDFs) in a fast and robust way. Its format is based on the standard $\texttt{LHEFv3}$ format~\cite{Alwall:2006yp}. The existence of this tool saves both computational time and disk space, but its most profound feature  is that it supplies analysers with  a weighted sample to account for the various theoretical systematics with the same exact size and the same statistical power (S.P.) as the one of the nominal/central sample. 
\\

To support the last statement, and although it may be trivial to the more advanced user, for the sake of completeness we consider the simple case where the central sample  consists of $N$ total events events and each events has a weight of $1$ ($\mathcal{W} = 1$). Then, we define the S.P. of the central sample as:

\begin{equation}
\label{weight}
S.P. = \frac{ \sum\limits_{n=1}^{N}  {\mathcal{W}}} { \sqrt{\sum\limits_{n=1}^{N}{\mathcal{W}^2}}} = \frac{ \sum\limits_{n=1}^{N}  {1}} { \sqrt{\sum\limits_{n=1}^{N}{1^2}}} = \frac{N}{\sqrt{N} }= \sqrt{N}
\end{equation}
\\

Assume now, that for the weighted sample, each event has a new weight  $\mathcal{W'}=\alpha$, and for simplicity lets assume that it is the same weight for all events. Similarly to Eq.~\ref{weight} we can write:

\begin{equation}\label{weight2}
S.P. = \frac{ \sum\limits_{n=1}^{N}  {\mathcal{W'}}} { \sqrt{\sum\limits_{n=1}^{N}{\mathcal{W'}^2}}} = \frac{ \sum\limits_{n=1}^{N}  {\alpha}} { \sqrt{\sum\limits_{n=1}^{N}{\alpha^2}}} = \frac{\alpha N}{\sqrt{\alpha^2 N} }= \sqrt{N}
\end{equation}

It is obvious from the above example, that in the case when the dedicated systematic samples are $n$ times smaller than the nominal sample, the statistical power of the dedicated samples is decreased by a factor of $\sqrt{n}$.
 With the use of \syscalc, as the number of events for a given theoretical uncertainty variation is equal to the central one, the statistical uncertainty on the related systematics are kept at the same level as the ones of the central sample. 
\\

The structure of this paper is as follows: In Sect.~\ref{syscalc} we  briefly present the package, while in Sect.~\ref{unmatched} (\ref{matched}) we describe the derivation of the formulas used in order to calculate the various systematics uncertainties for the case of un-matched/un-merged (matched/merged) samples followed by some validation plots. Further, in Sect.~\ref{guide}  we give a practical guide of how to use the code and the paper closes with the conclusions in Sect.~\ref{close}.  In the generation of all plots in this paper, the \delphes~\cite{delphes} and \texttt{ExRootAnalysis} packages have been used.

  %!TEX root = SysCalc_paper.tex

\section{The \syscalc\ package}\label{syscalc}

\syscalc\ is a package that can calculate dedicated event weights for certain theoretical systematical uncertainties. 
\syscalc\  supports all Leading-Order (LO) computations generated in \mgamc~\cite{Alwall:2014hca,Alwall:2007fs}.
Its output is an XML-based file which contains all relative weights needed to account for the selected systematics.
Supported systematics include the variations of the  $\mu_{\rm{F}}$, $\mu_{\rm{R}}$ and $\alpha_s$ scales, as well as PDF sets and grids. 
\syscalc\ makes use of additional information stored by \mgamc\ inside the record for each event, providing access to all 
information required to recompute the event weight based on convolution of the PDF set with the Matrix Element (ME) for the various supported scale variations.
\clearpage

\section{The un-matched case}
\label{unmatched}

\subsection{Reweighing of the \texorpdfstring{$\mu_{\rm{F}},\mu_{\rm R}$} scales}

Without matching/merging, \syscalc\ is able to compute the variation of $\mu_{\rm{F}}$ and $\mu_{\rm{R}}$ scales (parameter  \emph{scalefact}) and the change of the PDF set.
The variation of the scales can be done in a correlated and/or a uncorrelated way which is controlled by the value of the \textit{scale-correlation} parameter which can take the following values:
\begin{itemize}
\item  -1: to account for all combinations ($N^2$).
\item  -2: to account only for the correlated variations.
\item A set of positive values corresponding to the following entries (assuming \emph{0.5, 1, 2} for the  \emph{scalefact} entry):
\begin{enumerate}
\item[0:] $\mu_{\rm{F}} =\mu^{orig}_{\rm{F}} /2,~~~ \mu_{\rm{R}} =\mu^{orig}_{\rm{R}} /2$
\item[1:] $\mu_{\rm{F}} =\mu^{orig}_{\rm{F}} /2,  ~~~\mu_{\rm{R}} =\mu^{orig}_{\rm{R}} $
\item[2:] $\mu_{\rm{F}} =\mu^{orig}_{\rm{F}} /2,~~~\mu_{\rm{R}} =\mu^{orig}_{\rm{R}} *2$
\item[3:] $\mu_{\rm{F}} =\mu^{orig}_{\rm{F}},~~~\mu_{\rm{R}} =\mu^{orig}_{\rm{R}} /2$
\item[4:] $\mu_{\rm{F}} =\mu^{orig}_{\rm{F}},  ~~~\mu_{\rm{R}} =\mu^{orig}_{\rm{R}} $
\item[5:] $\mu_{\rm{F}} =\mu^{orig}_{\rm{F}},  ~~~\mu_{\rm{R}} =\mu^{orig}_{\rm{R}} *2$
\item[6:] $\mu_{\rm{F}} =\mu^{orig}_{\rm{F}} *2, ~~~\mu_{\rm{R}} =\mu^{orig}_{\rm{R}} /2$
\item[7:] $\mu_{\rm{F}} =\mu^{orig}_{\rm{F}} *2, ~~~\mu_{\rm{R}} =\mu^{orig}_{\rm{R}}  $
\item[8:] $\mu_{\rm{F}} =\mu^{orig}_{\rm{F}} *2,~~~\mu_{\rm{R}} =\mu^{orig}_{\rm{R}} *2$
\end{enumerate}
\end{itemize}

The weight associated with the renormalisation scale is the following:

\begin{equation}
\mathcal{W^{\mu_{\rm{R}}}_{\rm {new}}} =  \frac{\alpha_S^{N}(\rm{\Delta*\mu_{\rm{R}}})}{\alpha^{N}_S(\rm{\mu_{\rm{R}}})}  *\mathcal{W_{\rm{orig}}}, 
\end{equation}
\\
where $\Delta$ is the scale variation considered, $\mathcal{W_{\rm{orig}}}$ and $\mathcal{W_{\rm{new}}}$ are the original and new weights associated to the 
event respectively. $N$ is the power in the strong coupling for the associated event (interferences are not taken account in a event by event basis).
\\
The weight associated to the scaling of the factorisation scale is:

\begin{equation}
\mathcal{W^{\mu_{\rm{F}}}_{\rm {new}}} =   \frac{f_{1,orig} (x_1, \Delta*\mu_{\rm{F}}) * f_{2,orig} (x_2, \Delta*\mu_{\rm{F}})} {f_{1,orig}(x_1, \mu_{\rm{F}}) * f_{2,orig}(x_2, \mu_{\rm{F}})} *\mathcal{W_{\rm{orig}}}, 
\end{equation}
\\
where $f_{i,orig}$ are the PDF sets associated to the particle (which holds a fraction of energy $x_1$/$x_2$ for the first/second beam respectively) for the original PDF set.\\
% The computation is done for a dynamical scale ($\mu^0=H_T/2$)

As a validation, two typical processes, namely $ p p  \to t \bar t ~+xj$ and $ p p \to Z ~+xj, ~Z \to l^+ l^-$, with $x=0,1$ have been generated with \mg.
Around 10Mi events were produced for the central sample with a choice of $ \mu_{\rm R} = \mu_{\rm F} = \mu^{0}=H_{\rm {T/2}}$ (where $H_{\rm T}$ is the scalar sum of transverse 
momenta ({\pt}) of all jets of the event). The sample is then interfaced to
 \syscalc\ to derive the weights for the different scales into question. For each of the aforementioned processes, a set of dedicated samples is generated as well with 
 equal statistics, one for each of the following ($\mu_{\rm F}, \mu_{\rm R}$) configurations: 
\begin{itemize}
\item  $\mu_{\rm{R}} = 0.5\times\mu_0, ~\mu_{\rm{F}} = 1 \times \mu_0$
\item  $\mu_{\rm{R}} = 1 \times \mu_0, ~\mu_{\rm{F}} = 0.5 \times \mu_0$
\item  $\mu_{\rm{R}} = 1 \times \mu_0, ~\mu_{\rm{F}} = 2 \times \mu_0$
\item  $\mu_{\rm{R}} = 2 \times \mu_0, ~\mu_{\rm{F}} = 1 \times \mu_0$
\end{itemize}

\begin{figure}[!htp]
\vspace{-1cm}
 \captionsetup[subfigure]{labelformat=empty}
    \centering
    \subfloat[]{
       \begin{minipage}{\linewidth}

           \includegraphics[width=0.5\linewidth, height = 0.3\textheight, keepaspectratio=true]{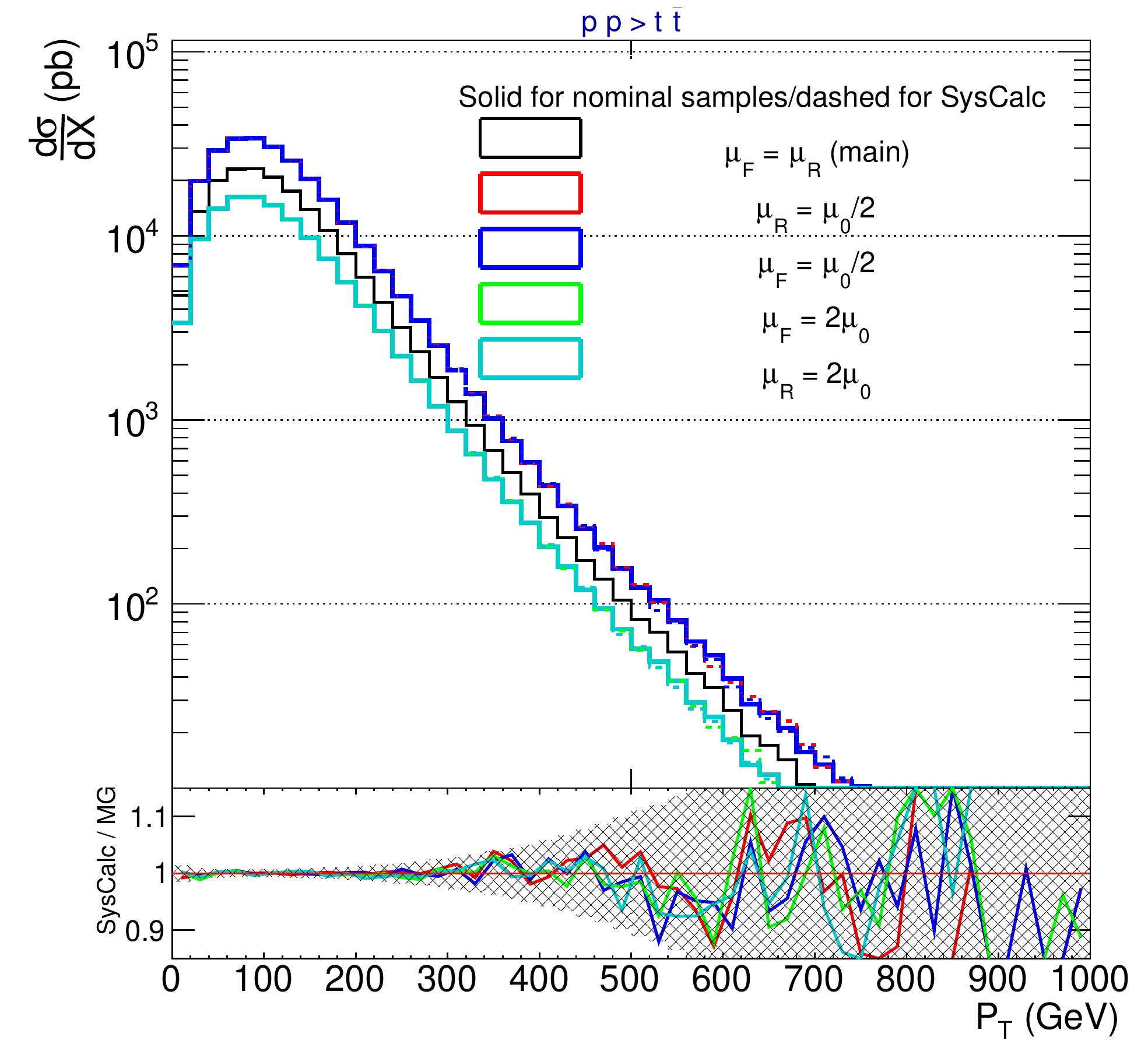}
            \includegraphics[width=0.5\linewidth, height = 0.3\textheight, keepaspectratio=true]{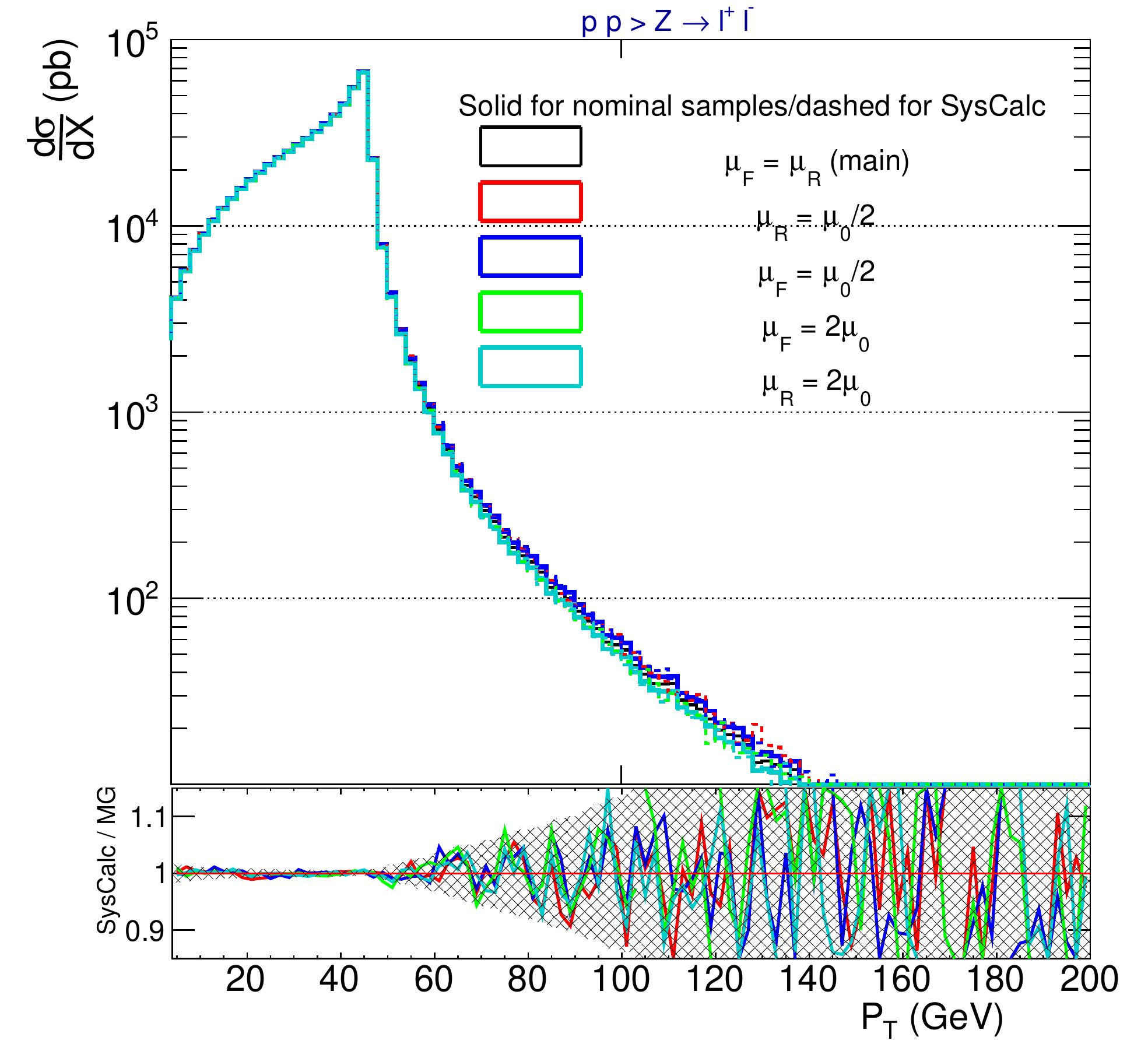}

           \includegraphics[width=0.5\linewidth, height = 0.3\textheight, keepaspectratio=true]{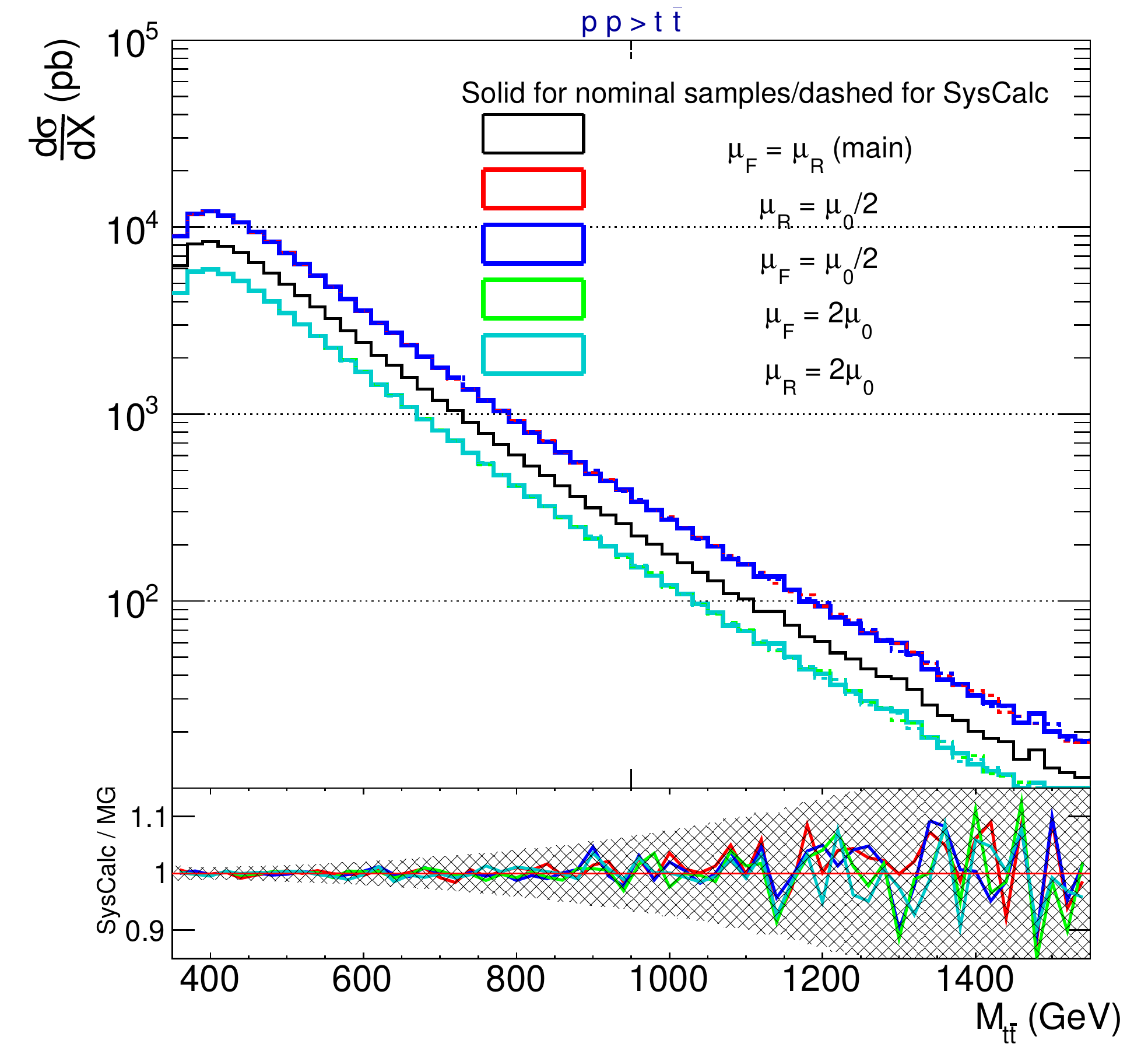}
            \includegraphics[width=0.5\linewidth, height = 0.3\textheight, keepaspectratio=true]{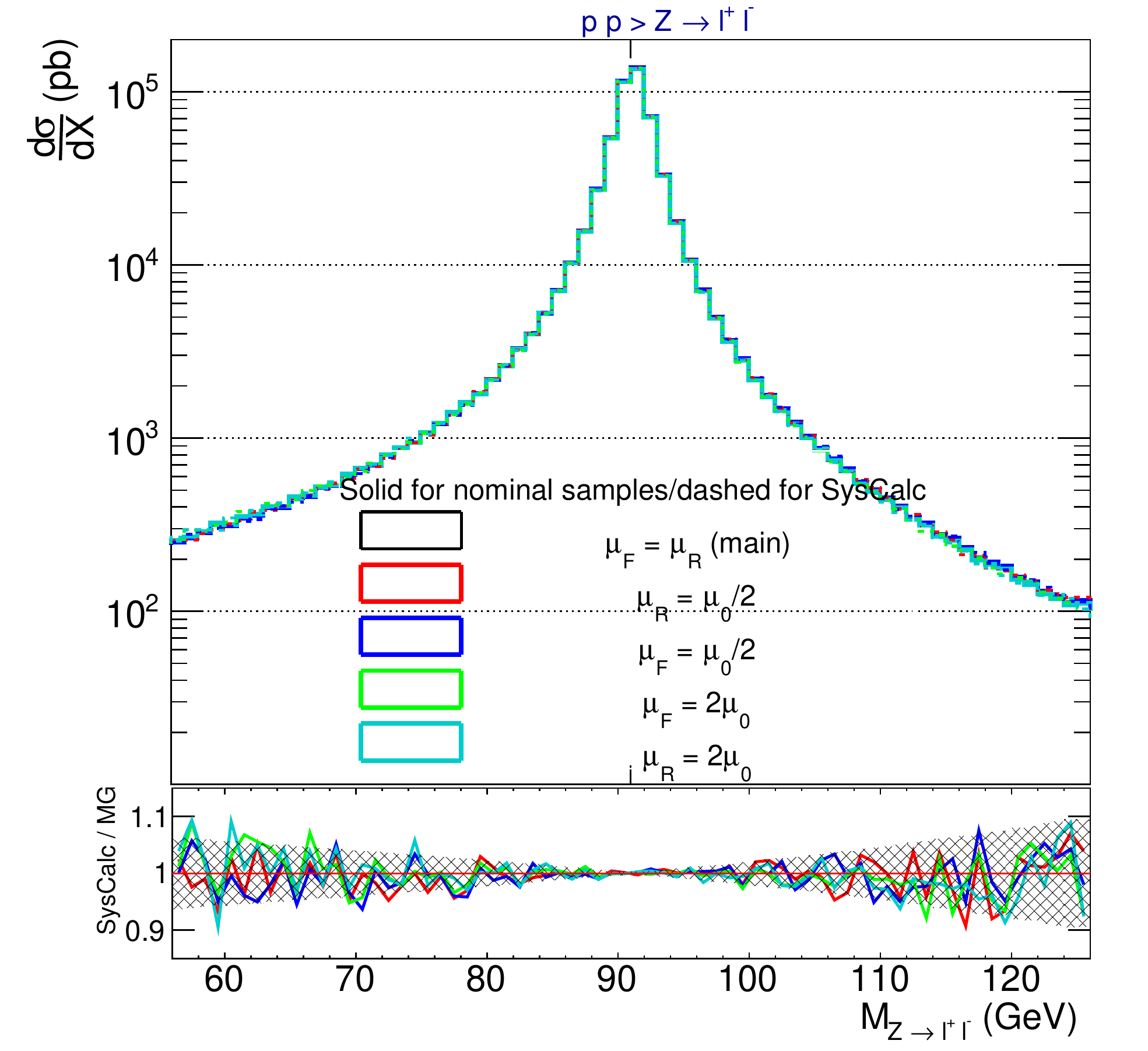}
            
     	 \includegraphics[width=0.5\linewidth, height = 0.3\textheight, keepaspectratio=true]{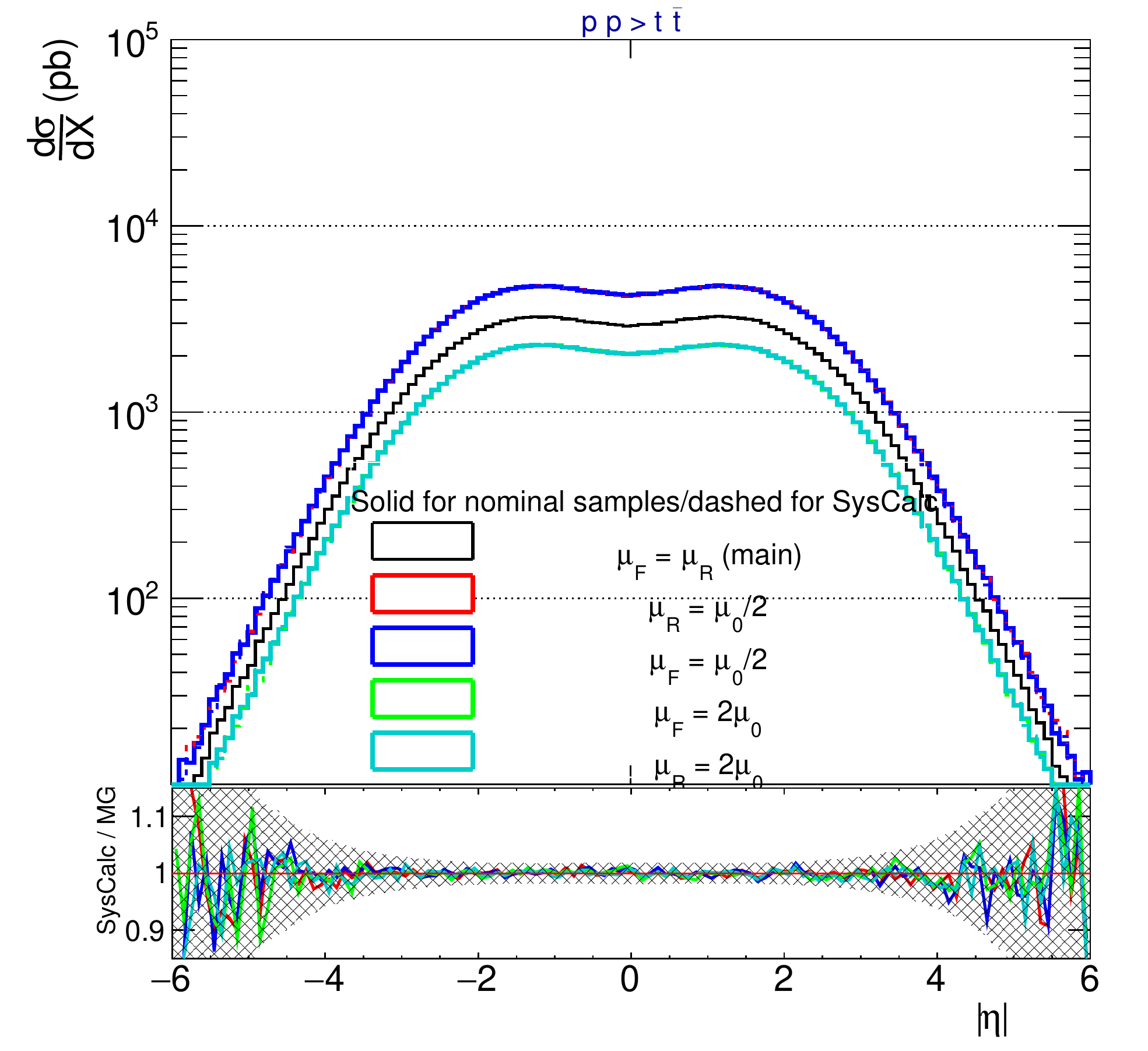}
          \includegraphics[width=0.5\linewidth, height = 0.3\textheight, keepaspectratio=true]{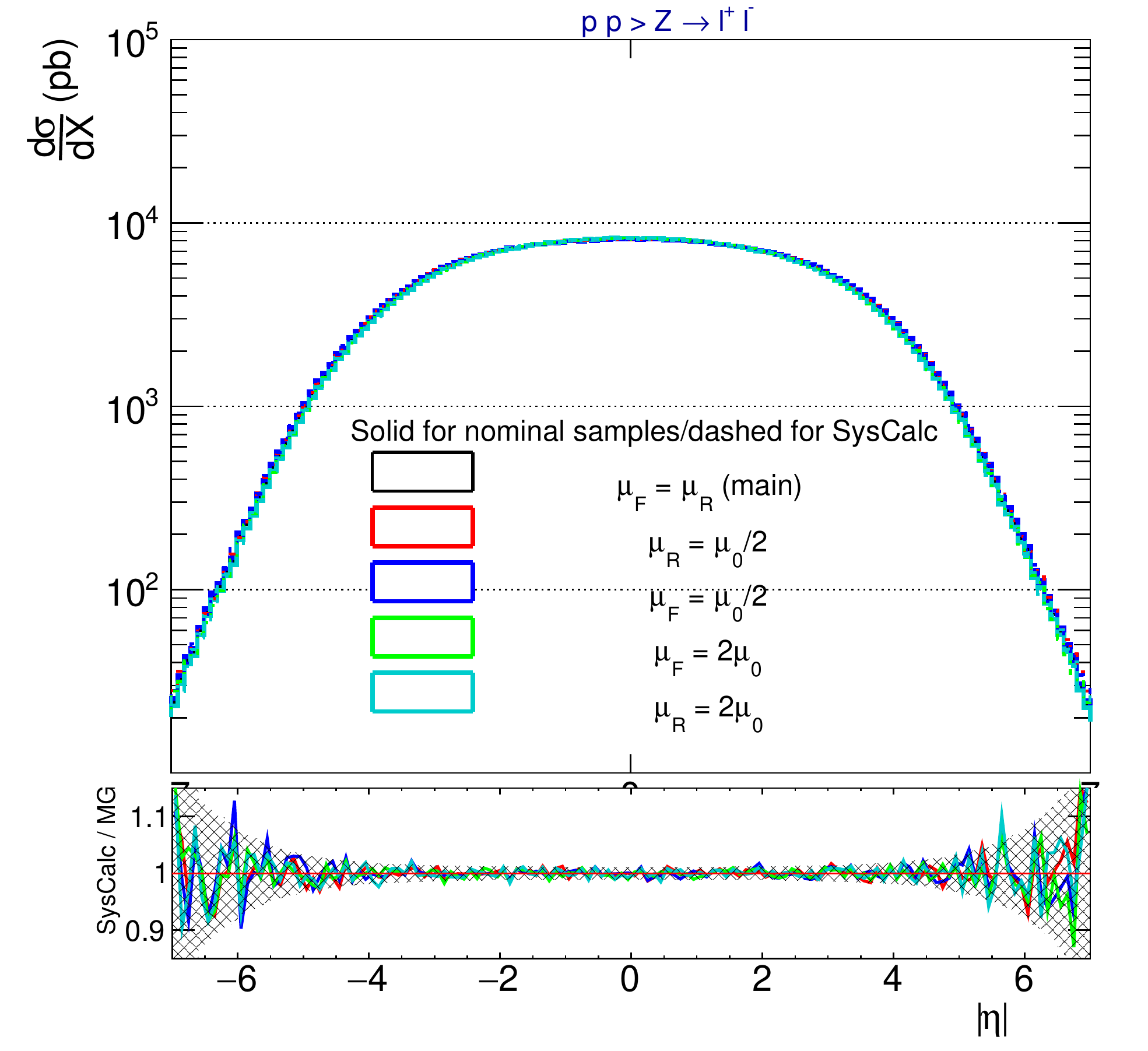}
             
             \end{minipage}}
             
 \caption{Comparison of the invariant mass, the \pt\ and the $|\eta|$ distributions of the $ p p  \to t \bar t ~+xj$ (left column) and $ p p \to Z ~+xj, ~Z \to l^+ l^-$ (right column) processes, with $x=0,1$, and between dedicated samples  for various choices of the $\mu_{\rm F}, \mu_{\rm R}$  scales, and the weighted events from \syscalc\ derived from the central sample (black solid line). The ratio between the dedicated samples and the weighted events is also shown at the lower canvas of each plot. The shaded area corresponds to the statistical uncertainty from the central sample.
 }
  \label{fig:scales_unmatched} 

 \end{figure}

For all of the generated samples, a unique random seed is initialised every 100k events. In Fig.~\ref{fig:scales_unmatched}, the invariant mass, the \pt and the rapidity ($|\eta|$) distributions of the top quark pair and of the leptons from the Z-bosons from the dedicated samples with different choices of the various scales are compared against the \syscalc\ weighted events. The agreement  between them is found to be within the statistical fluctuations.

\subsection{Reweighing of the PDF set}
\label{pdf}
The equation describing both the variation of the PDF set and the central scale in the case of fixed-order samples is given by the corresponding weights associated to the chosen new PDF sets and new central scales:
\begin{equation}
\mathcal{W^{PDF}_{\rm new}} =  \frac{f_{1,new} (x_1, \mu_{\rm{F}}) * f_{2,new} (x_2, \mu_{\rm{F}})} {f_{1,orig}(x_1, \mu_{\rm{F}}) * f_{2,orig}(x_2, \mu_{\rm{F}})}*   \frac{\alpha_{S,new}^{N}(\mu_{\rm{R}}) }{\alpha_{S,orig}^{N}(\mu_{\rm{R}}) }  \mathcal{W_{\rm{orig}}}, 
\end{equation}
\newline

where $f_{i,new}$ are the PDF sets associated to the particle under consideration.
\\

As  validation, the $ p p  \to t \bar t ~xj$  and $ p p \to Z ~xj, Z \to l^+ l^-$ processes (with $x=0,1$) have been considered. The central sample has been generated with the NNPDF\_{LO}($\alpha_S = 0.13$) PDF, while the events have been weighted for the following PDF sets \cite{Lai:2010vv,Harland-Lang:2014zoa,Gao:2013xoa,Pumplin:2002vw,Ball:2014uwa}:

\begin{itemize}
\item  CT10nlo
\item MMHT2014nlo68cl
\item CT10nnnlo~${\alpha_S} = 0.130$
\item cteq6l1
\item MMHT2014lo68cl
\item NNPDF30nnlo~${\alpha_S} =0.118$
\end {itemize}

\begin{figure}[!htp]
\vspace{-1cm}

 \captionsetup[subfigure]{labelformat=empty}
    \centering
    \subfloat[]{
       \begin{minipage}{\linewidth}
            \includegraphics[width=0.5\linewidth, height = 0.3\textheight, keepaspectratio=true]{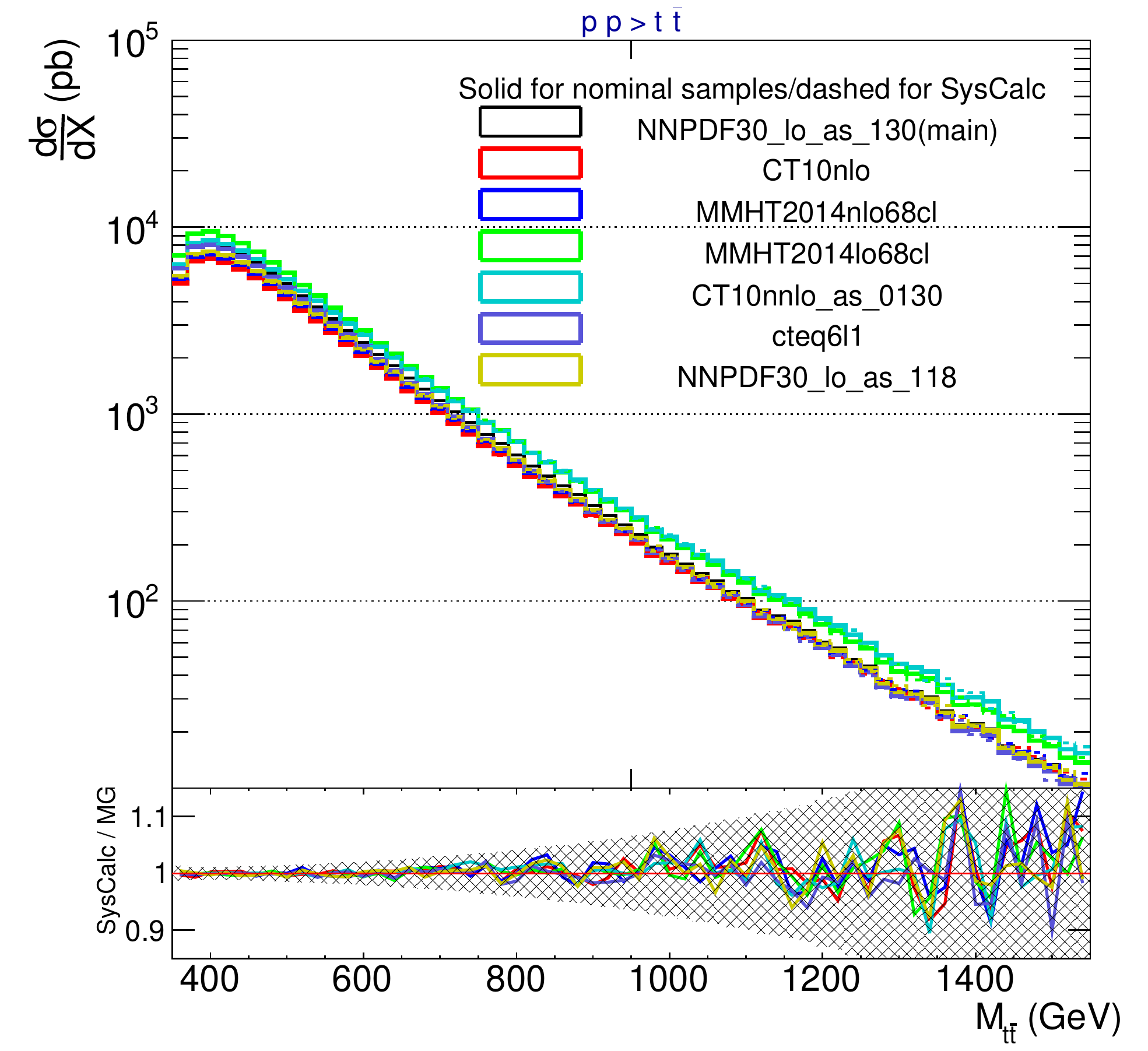}
            \includegraphics[width=0.5\linewidth, height = 0.3\textheight, keepaspectratio=true]{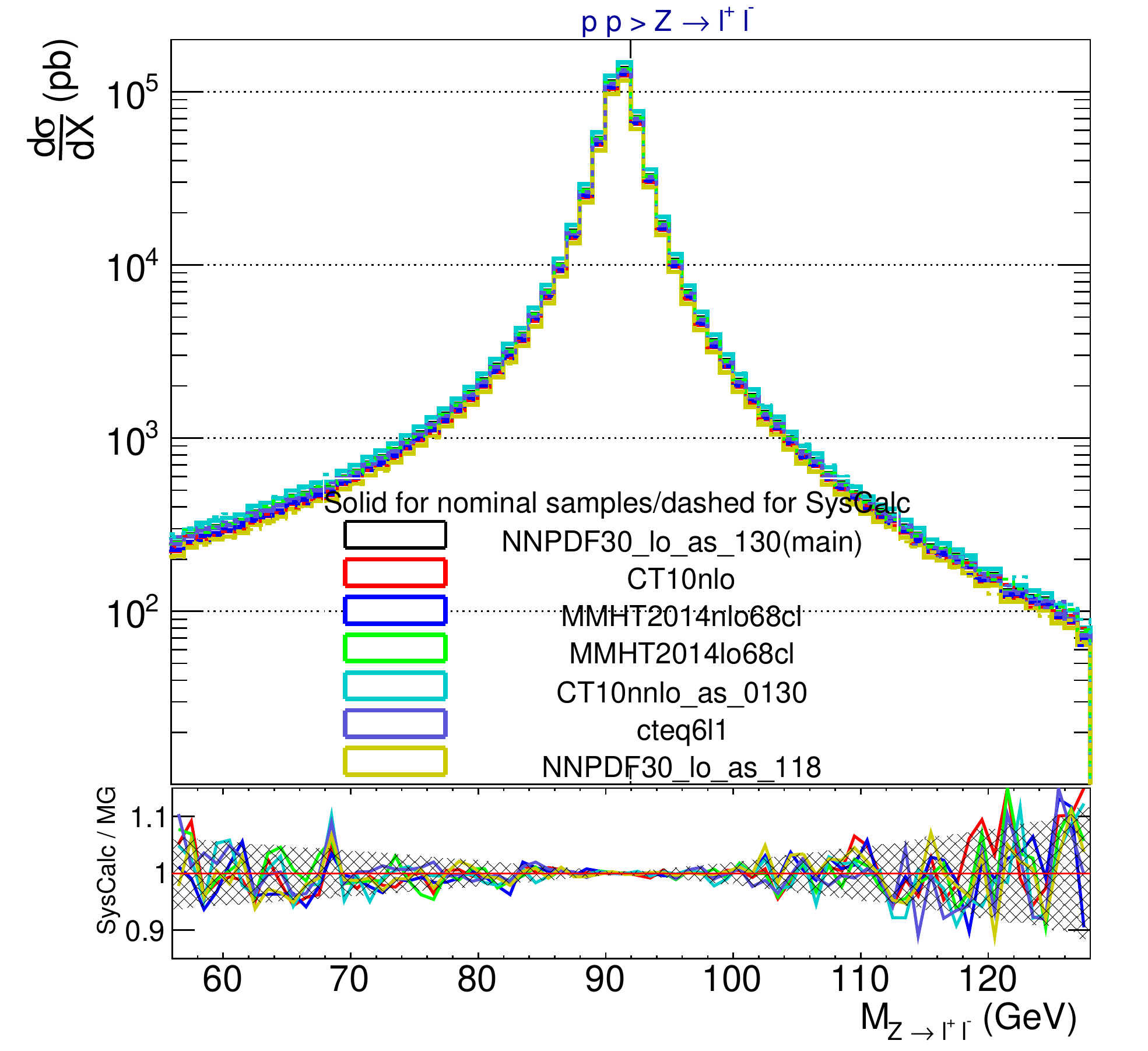}
            \\
            \includegraphics[width=0.5\linewidth, height = 0.3\textheight, keepaspectratio=true]{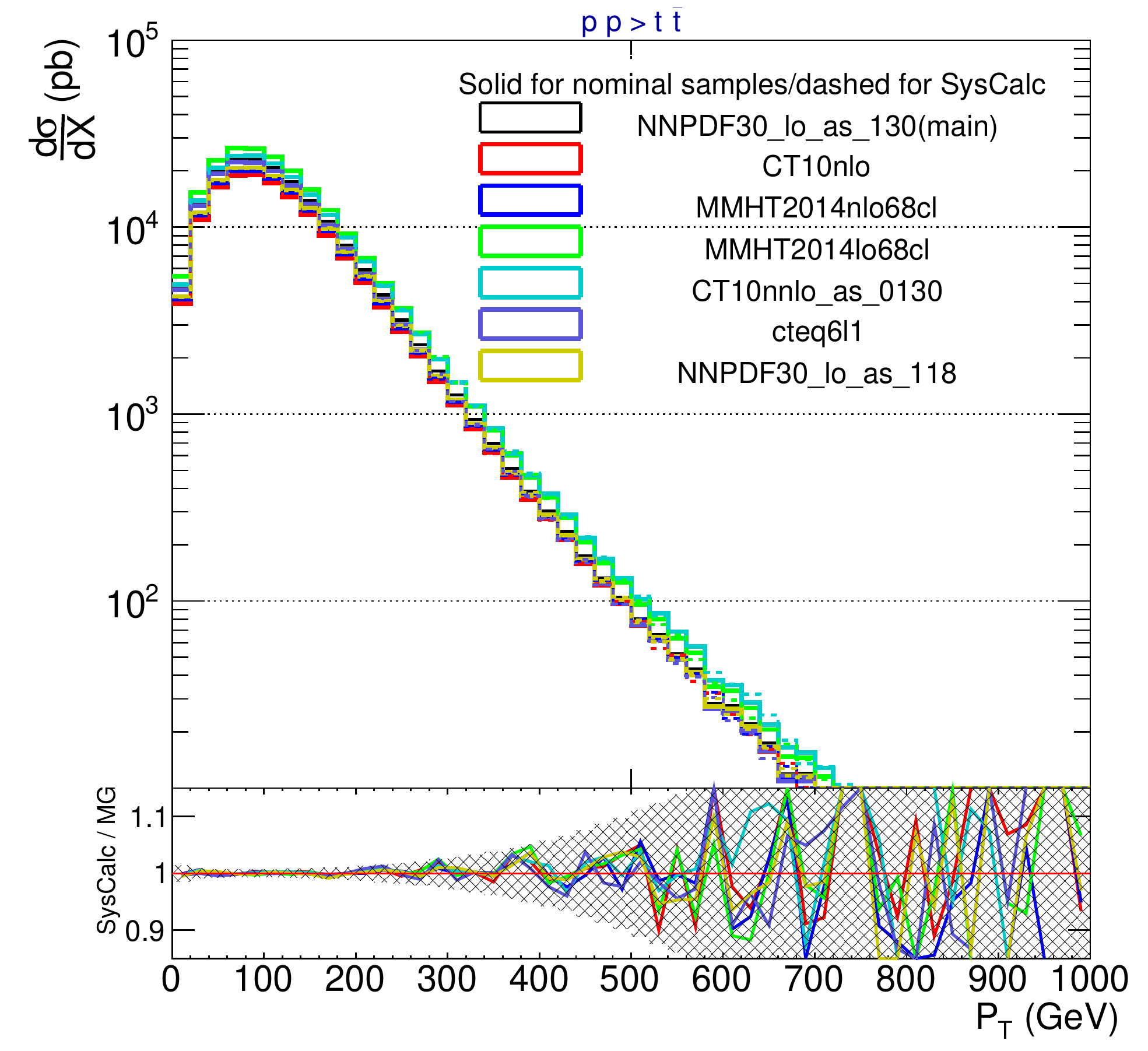}
            \includegraphics[width=0.5\linewidth, height = 0.3\textheight, keepaspectratio=true]{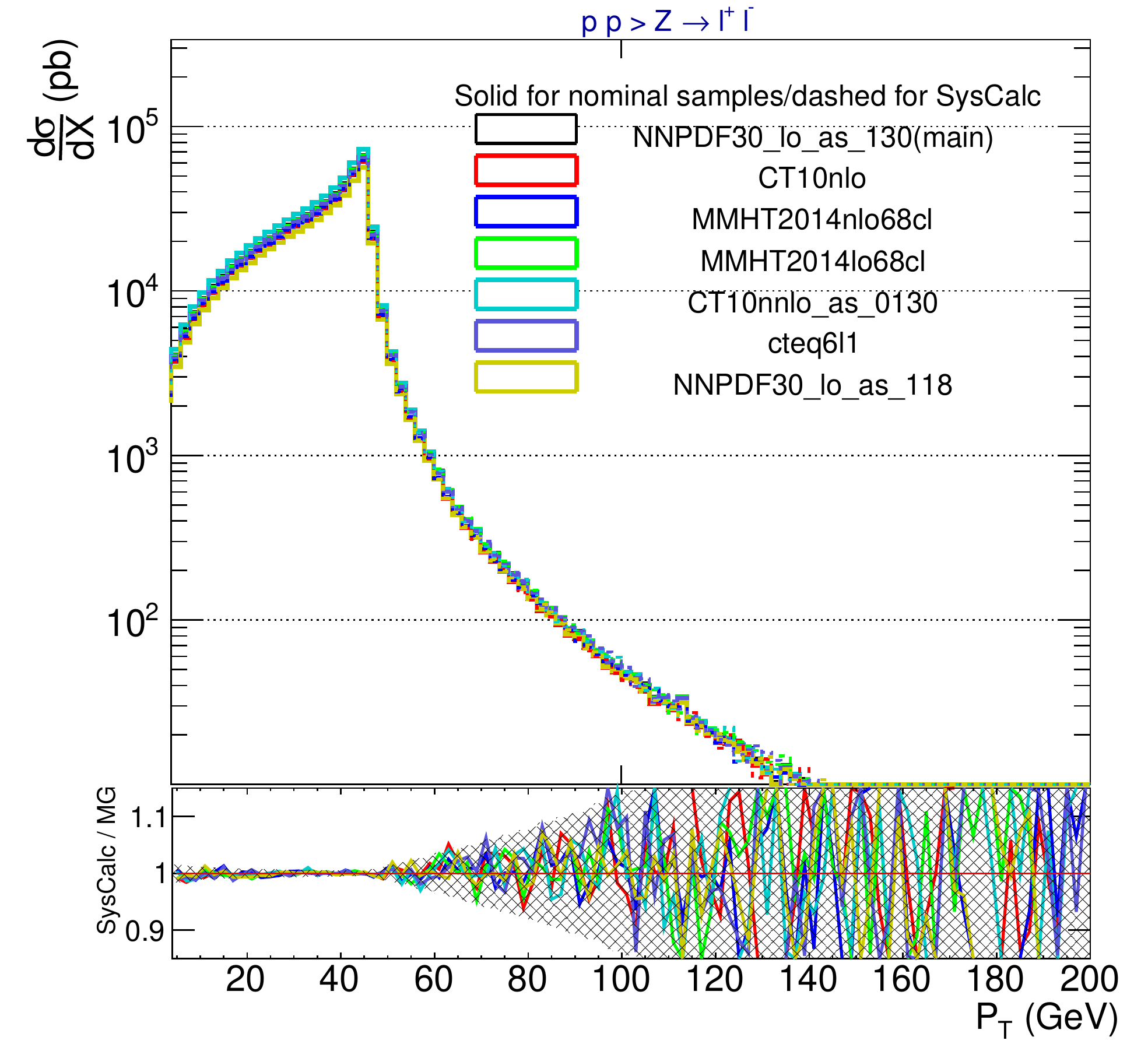}
            \\
            \includegraphics[width=0.5\linewidth, height = 0.3\textheight, keepaspectratio=true]{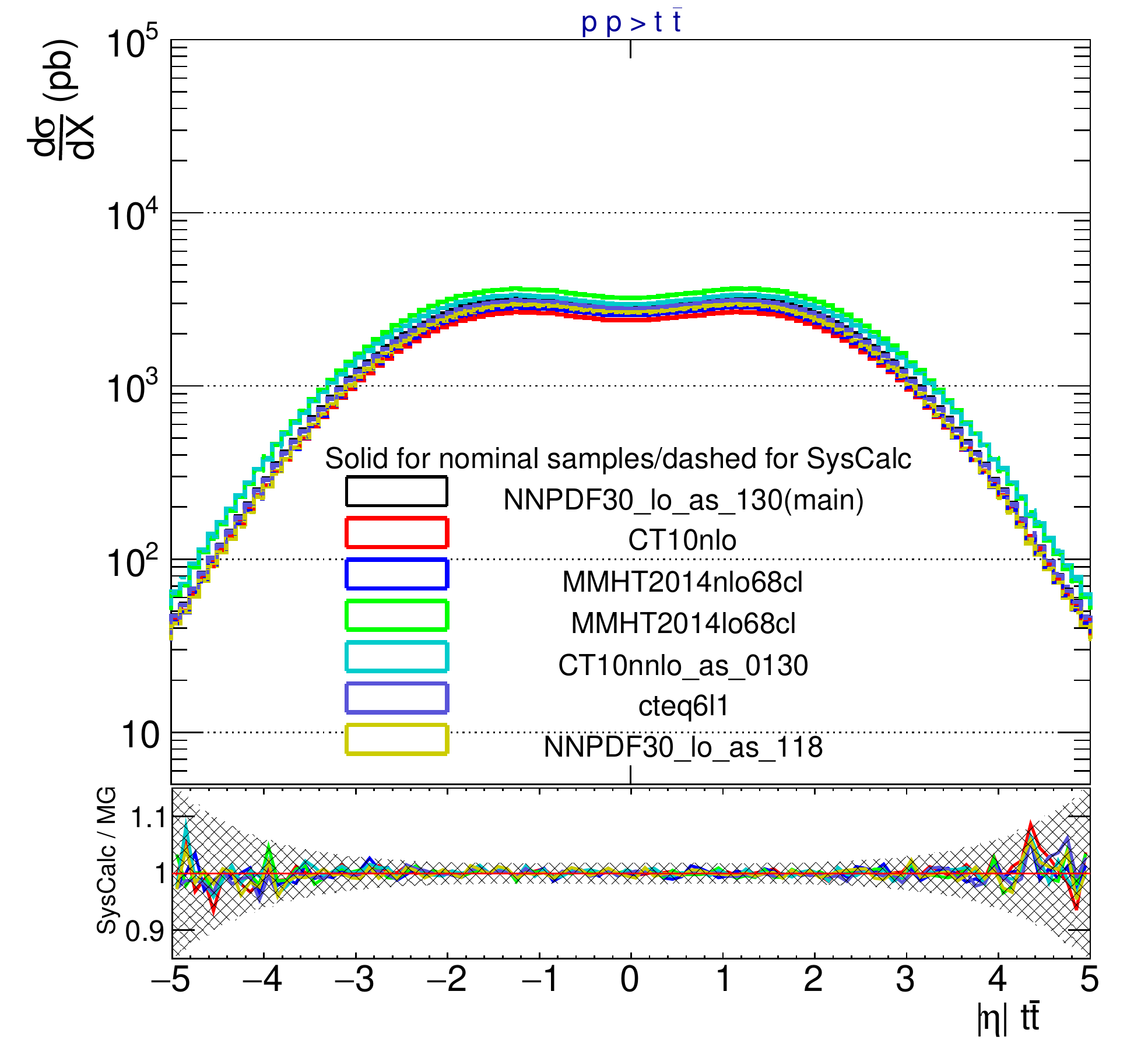}
            \includegraphics[width=0.5\linewidth, height = 0.3\textheight, keepaspectratio=true]{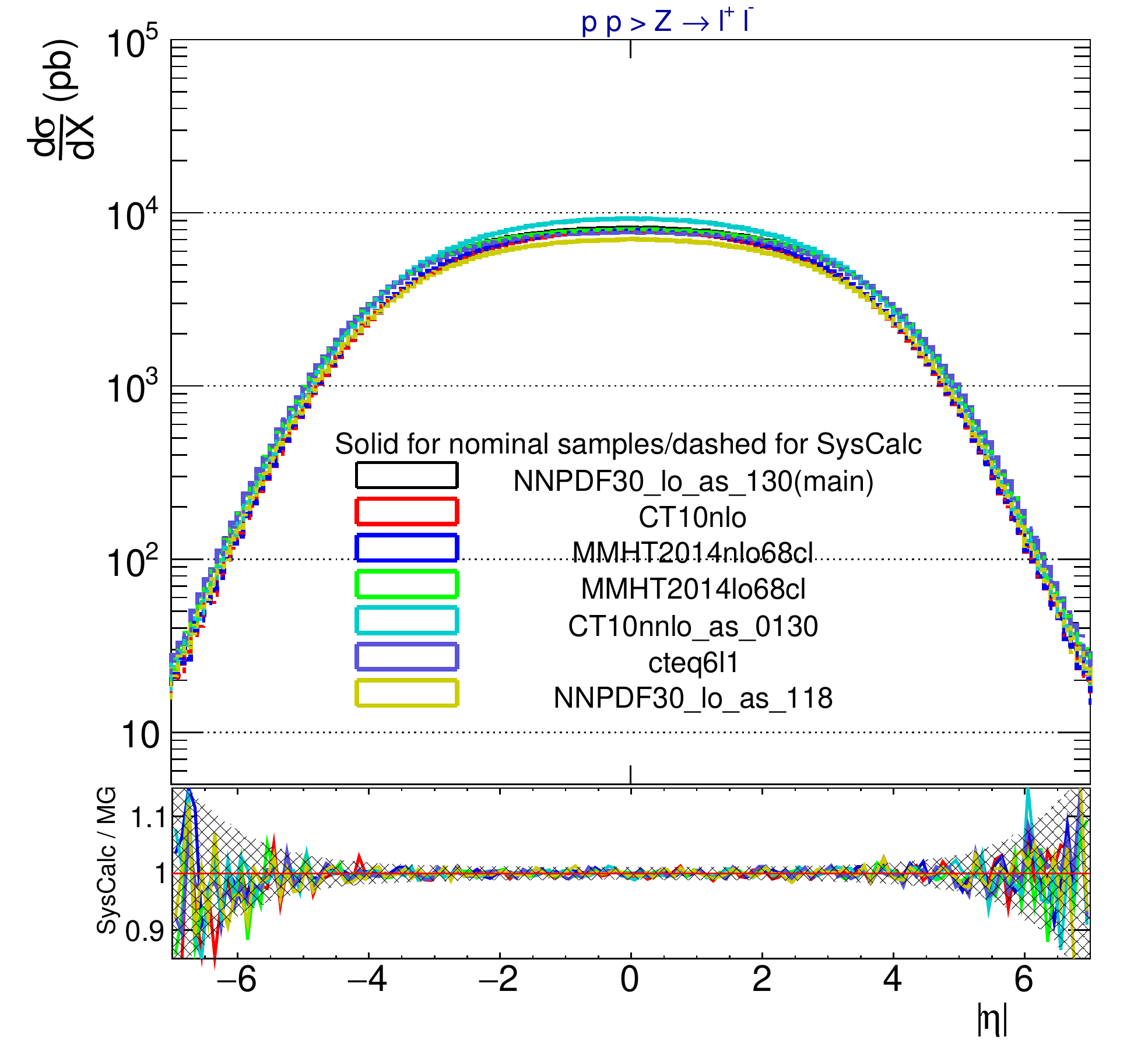}
            
             \end{minipage}}

 \caption{Comparison of the invariant mass, the \pt\ and the $|\eta|$ distributions of the $ p p  \to t \bar t ~+xj$ (left column) and $ p p \to Z ~+xj, ~Z \to l^+ l^-$ (right column) processes, with $x=0,1$,  between dedicated samples for various choices of the PDF set, and the weighted events from \syscalc\ derived from the central sample (black solid line). The ratio between the dedicated sample and the weighted events from is also shown at the bottom of each plot. The shaded area corresponds to the statistical uncertainty from the central sample.}
  \label{fig:pdf_unmathced} 

 \end{figure}

Further, dedicated samples have been generated with \mgamc\ for each of the above PDF sets for a choice of $\mu_{\rm{R}} = 1 = \mu_{\rm{F}} = 1 \times \mu_0$, while the central sample is generated with the NNPDF\_{LO}($\alpha_S = 0.13$) PDF set. 
In Fig.~\ref{fig:pdf_unmathced},  the invariant mass, the \pt, and the $|\eta|$ of the top quark pair and of the leptons from the Z-bosons from the dedicated samples from the different PDFs are compared against the \syscalc\ weighted events. The agreement  is found to be within the statistical fluctuations.

\clearpage
\section{The case of matching}
\label{matched}

In the presence of matching, \mgamc\ rescales each event, such that the scale of the strong interaction in emissions as well as the PDF is set according to the parton shower history
 (which is selected via a $k_T$ clustering). \syscalc\ can perform an associated re-weighting (parameter  $\emph{alpsfact}$) by dividing and by multiplying with the associated factor.
\\

For each vertex of the clustering (associated to a scale $\mu_i$) this corresponds to the following factor for Final State Radiation (FSR):
\begin{equation}
\mathcal{W^{FSR}_{\rm {new}}} = \frac{ \alpha_s (\Delta* \mu_i)} { \alpha_s (\mu_i)}  * \mathcal{W_{\rm{orig}}}
\end{equation}

and similarly for Initial State Radiation (ISR) associated to a scale $\mu_i$ and fraction of energy $x_i$:
\begin{equation}
\mathcal{W^{ISR}_{\rm new}} =\frac{ \alpha_s (\Delta* \mu_i)} { \alpha_s (\mu_i)} \frac{ \frac{f_a(x_i,\Delta*\mu_i)}{f_b(x_i,\Delta*\mu_{i+1})}} { \frac{f_a(x_i, \mu_i)}{f_b(x_i, \mu_{i+1})} }
  *\mathcal{W_{\rm{orig}}}
\end{equation}
where $\mu_{i+1}$ is the scale of the next vertex in the initial state clustering history.
\newline

To test \syscalc\ in the case of matching,  we have performed a similar validation as described in Sec.~\ref{unmatched} for different choices of the scales and different PDF sets as well. 
The utilized samples  have been generated with 0 and 1 parton at ME with \mg\ and with a different random seed every 100k events for about 20Mi events in total. The parton level events were interfaced with \pythia\cite{pythia8} for the hadronization and the matching step.\\
 
The validation was performed for different variations of the $\mu_{\rm{F}}, \mu_{\rm{R}},\rm{and~} \alpha_s$ scales:

\begin{itemize}
\item $\alpha_s' = \alpha_s \times 0.5$ 
\item $\alpha_s' = \alpha_s \times 2$ 
\item  $\mu_{\rm{F}}  = \mu_{\rm{R}} = 0.5\times\mu_0$
\item  $\mu_{\rm{F}}  = \mu_{\rm{R}} = 2\times\mu_0$
\end{itemize}

\begin{figure}[!htp]
\vspace{-1cm}
 \captionsetup[subfigure]{labelformat=empty}
    \centering
    \subfloat[]{
       \begin{minipage}{\linewidth}
        \includegraphics[width=0.5\linewidth, height = 0.3\textheight, keepaspectratio=true]{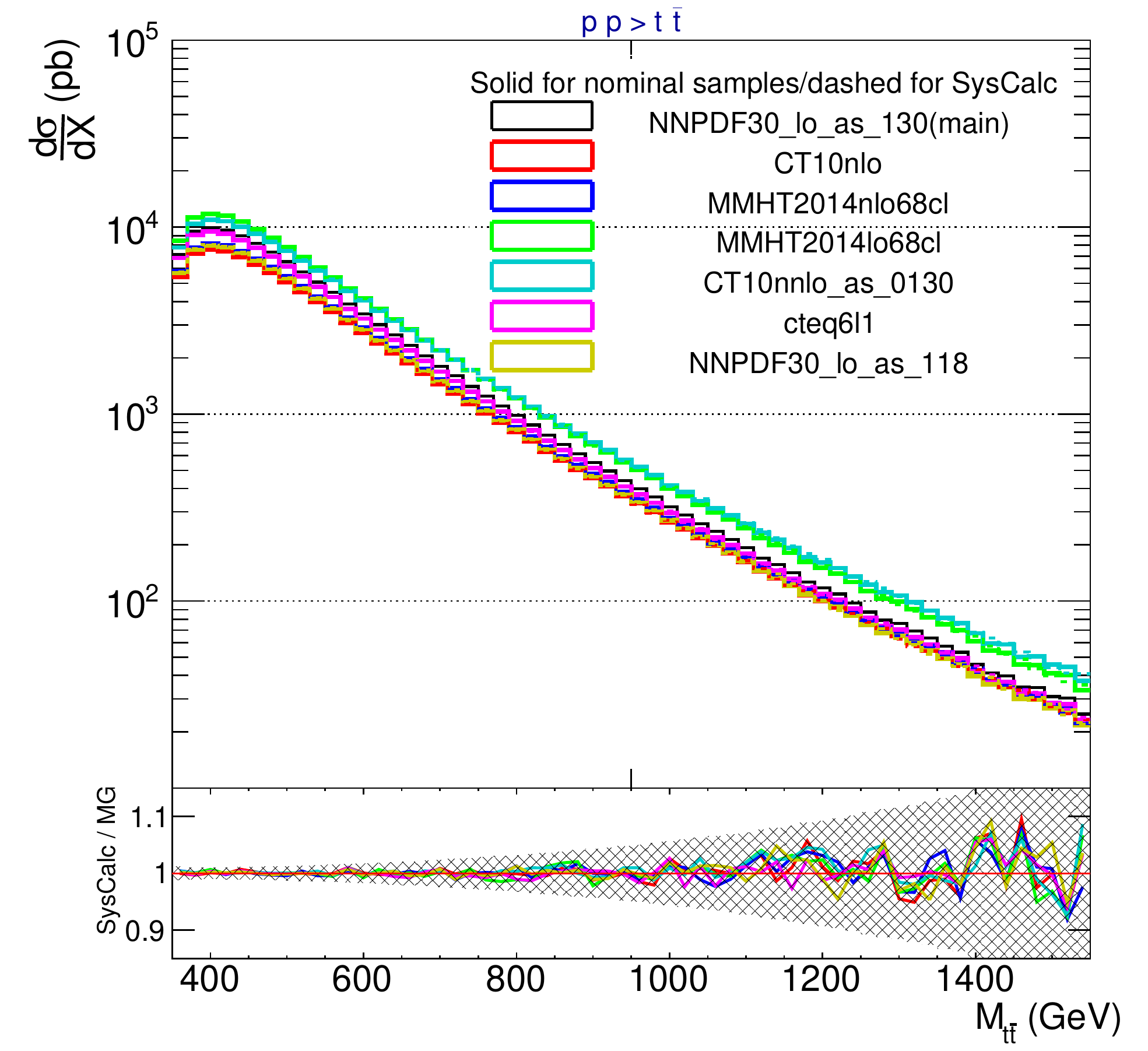}
            \includegraphics[width=0.5\linewidth, height = 0.3\textheight, keepaspectratio=true]{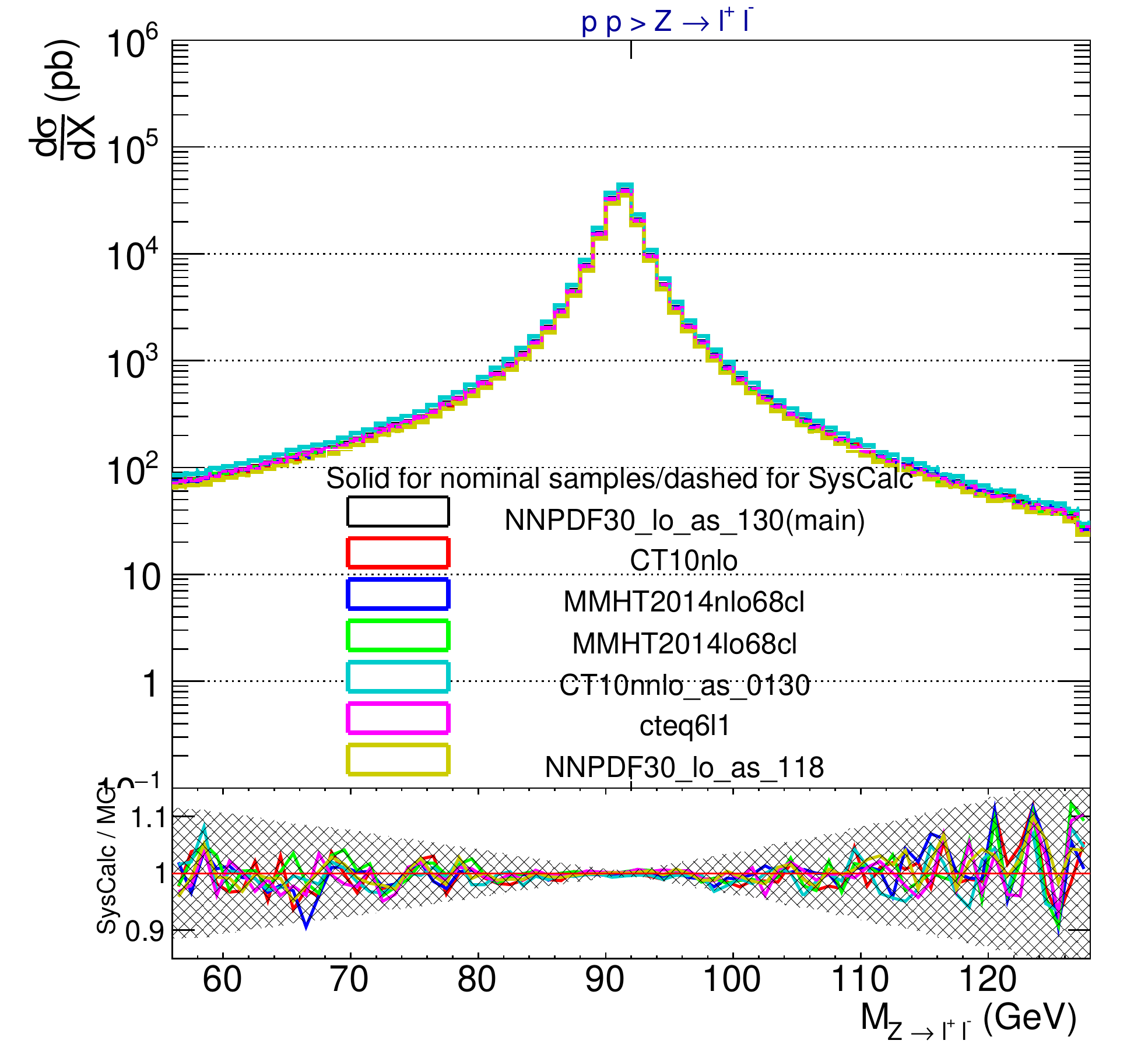}
     
          \includegraphics[width=0.5\linewidth, height = 0.3\textheight, keepaspectratio=true]{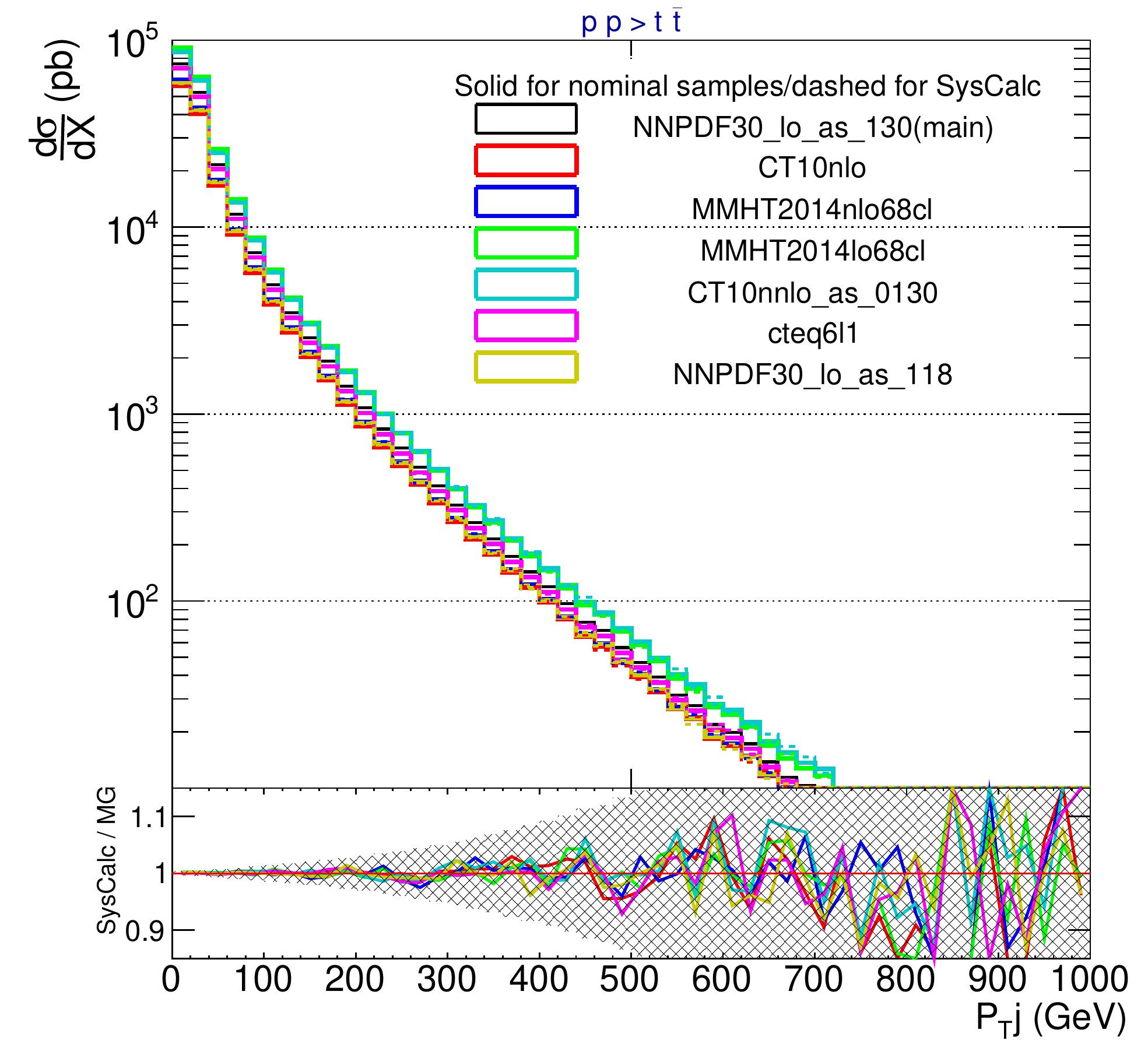}
          \includegraphics[width=0.5\linewidth, height = 0.3\textheight, keepaspectratio=true]{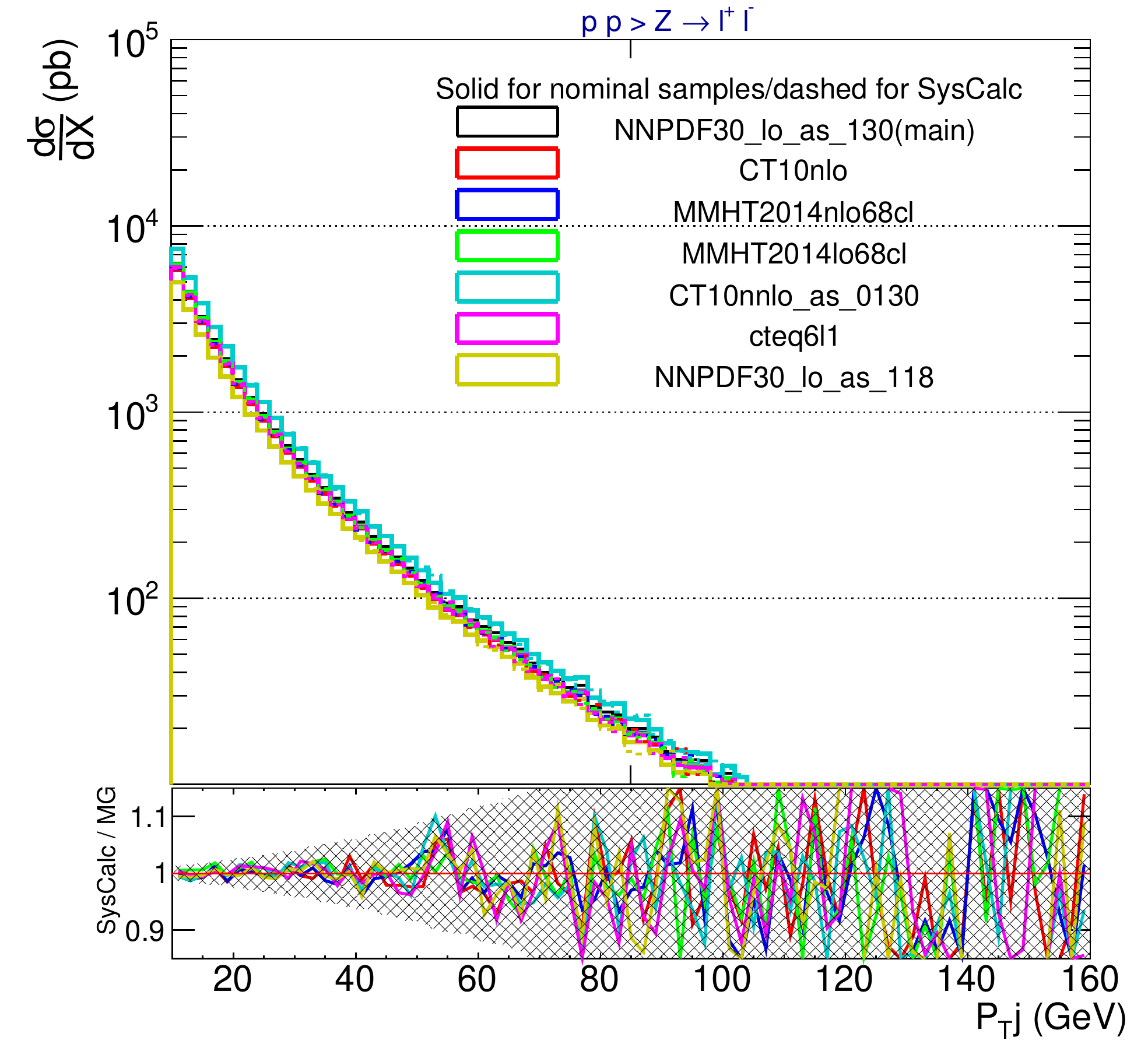}
    
          \includegraphics[width=0.5\linewidth, height = 0.3\textheight, keepaspectratio=true]{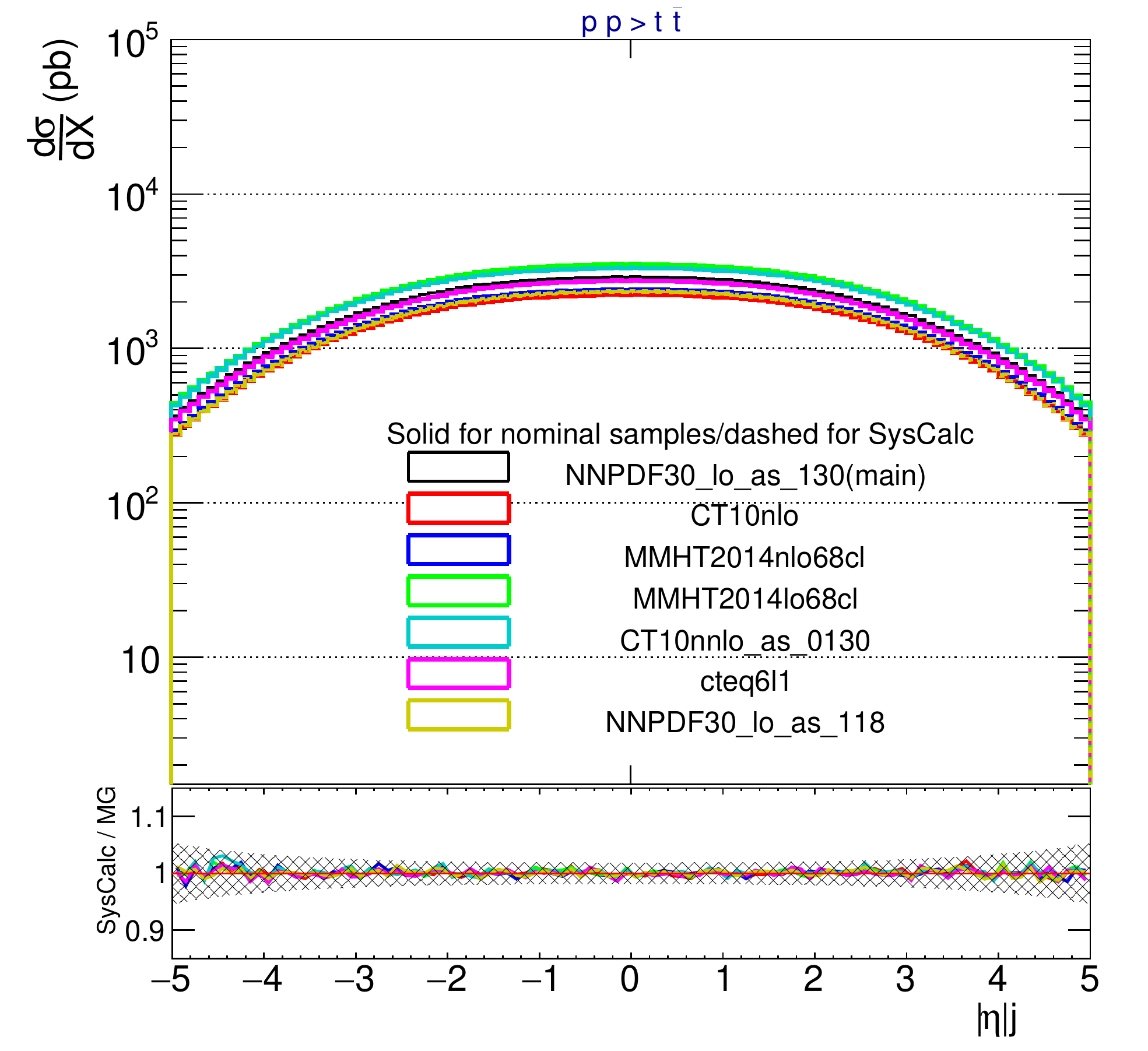}
          \includegraphics[width=0.5\linewidth, height = 0.3\textheight, keepaspectratio=true]{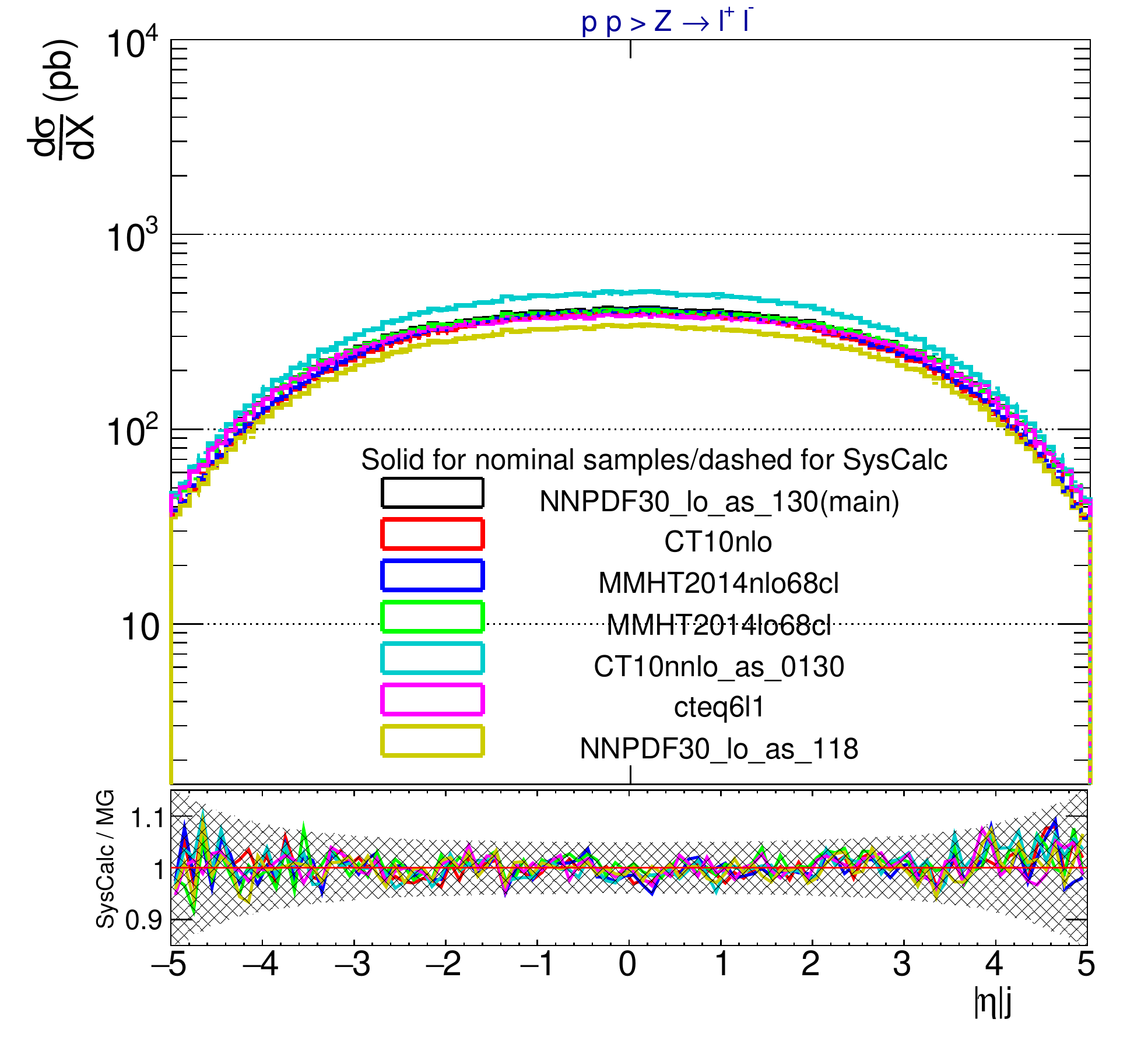}
      
             \end{minipage}}
             
 \caption{Distributions from $ p p  \to t \bar t ~+0,1j$ (left column) and $ p p \to Z ~+0,1j, ~Z \to l^+ l^-$ (right column) processes of the invariant mass  of the $t(\bar t)\rm{~and~the~}Z$-boson respectively, as well as the \pt\ and the $|\eta|$ distributions of the ISR jet for various different PDF sets, between the dedicated samples and the weighted events from \syscalc\ derived from the central sample (black solid line). The bottom on each distribution represents the ratio between the dedicated sample and the weighted events. The shaded area corresponds to the statistical uncertainty from the central sample.}
  \label{fig:pdf_matched} 

 \end{figure}

\begin{figure}[!htp]
\vspace{-2cm}
 \captionsetup[subfigure]{labelformat=empty}
    \centering
    \subfloat[]{
       \begin{minipage}{\linewidth}
        \includegraphics[width=0.5\linewidth, height = 0.3\textheight, keepaspectratio=true]{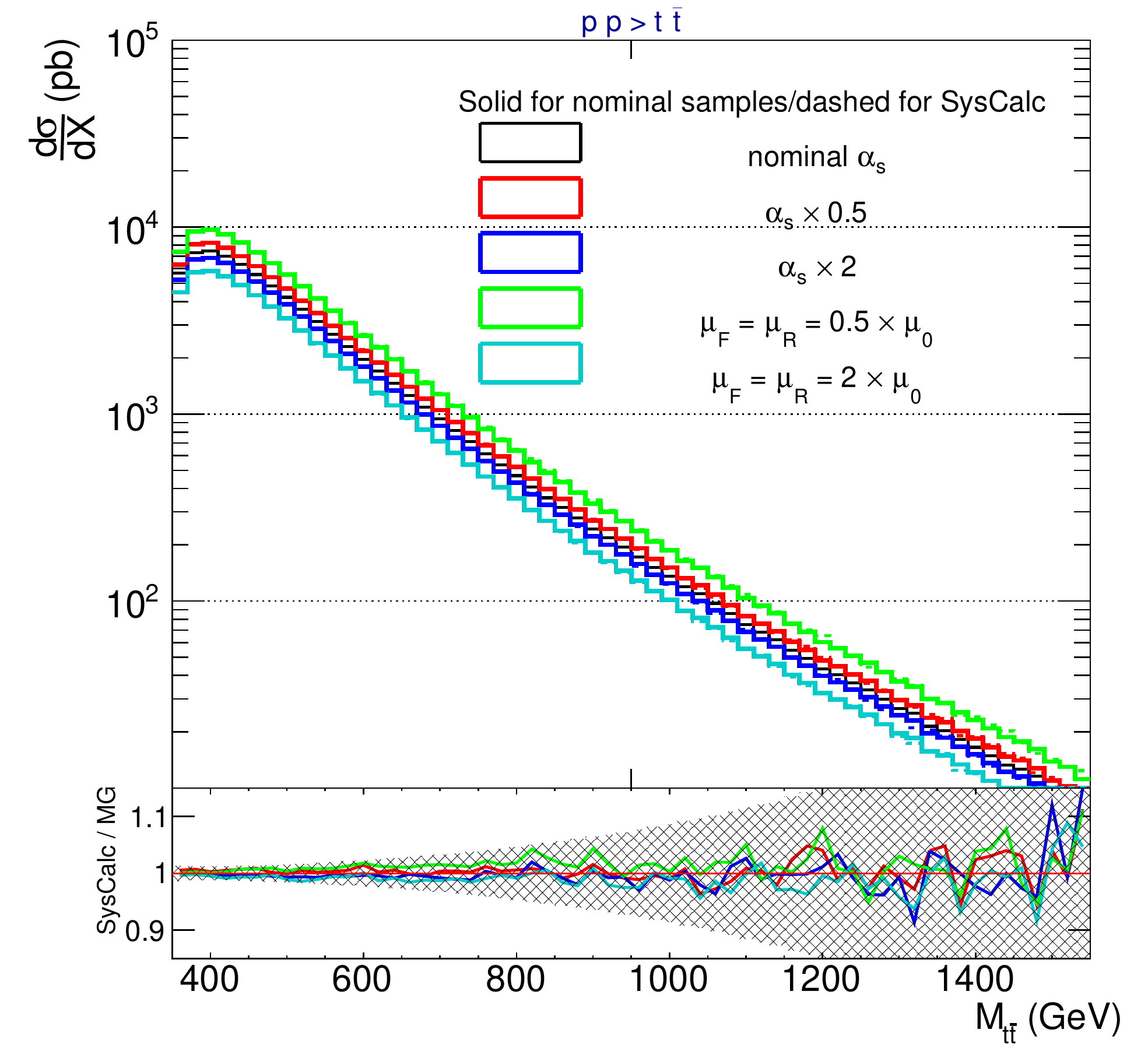}
            \includegraphics[width=0.5\linewidth, height = 0.3\textheight, keepaspectratio=true]{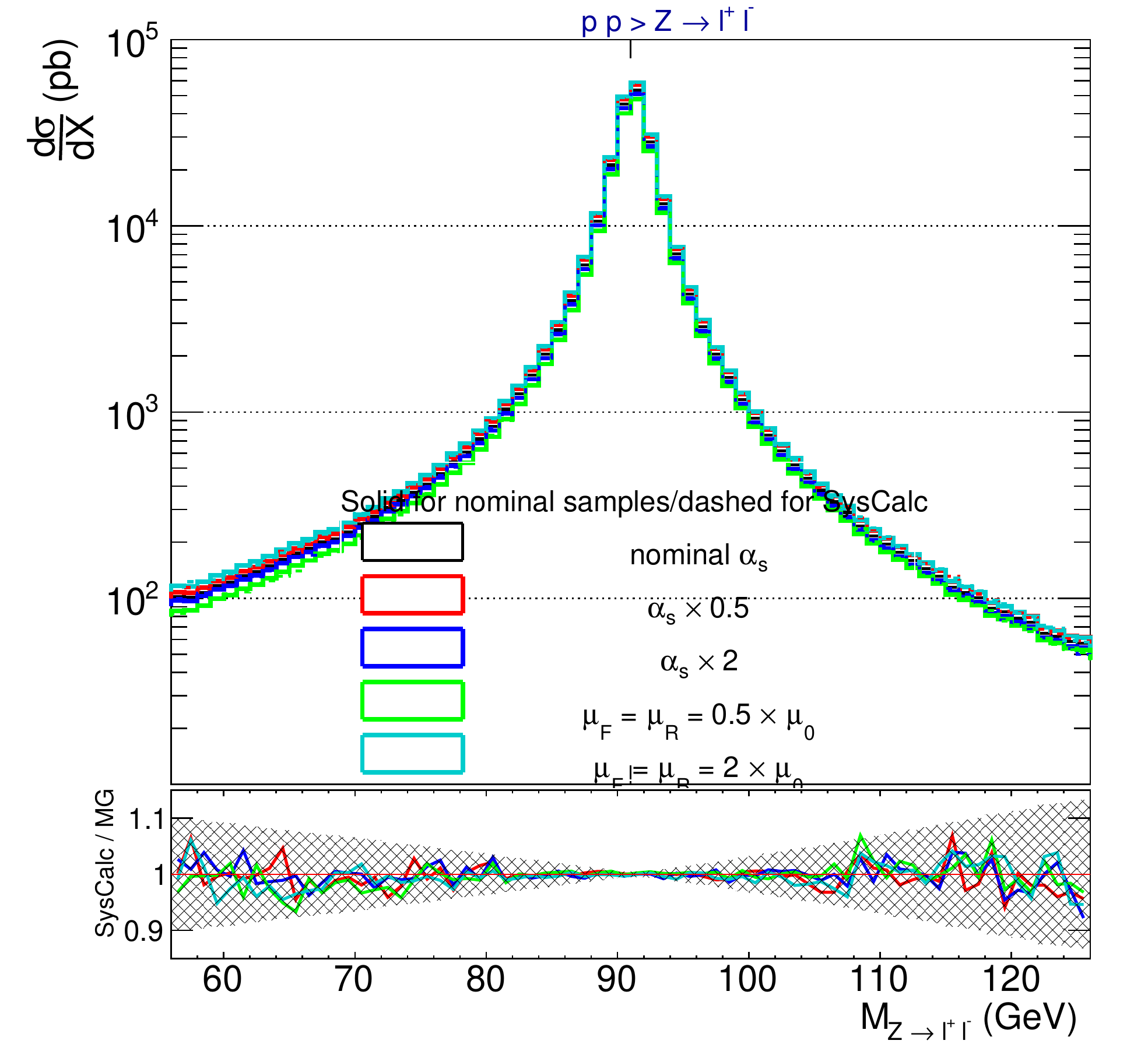}
     
          \includegraphics[width=0.5\linewidth, height = 0.3\textheight, keepaspectratio=true]{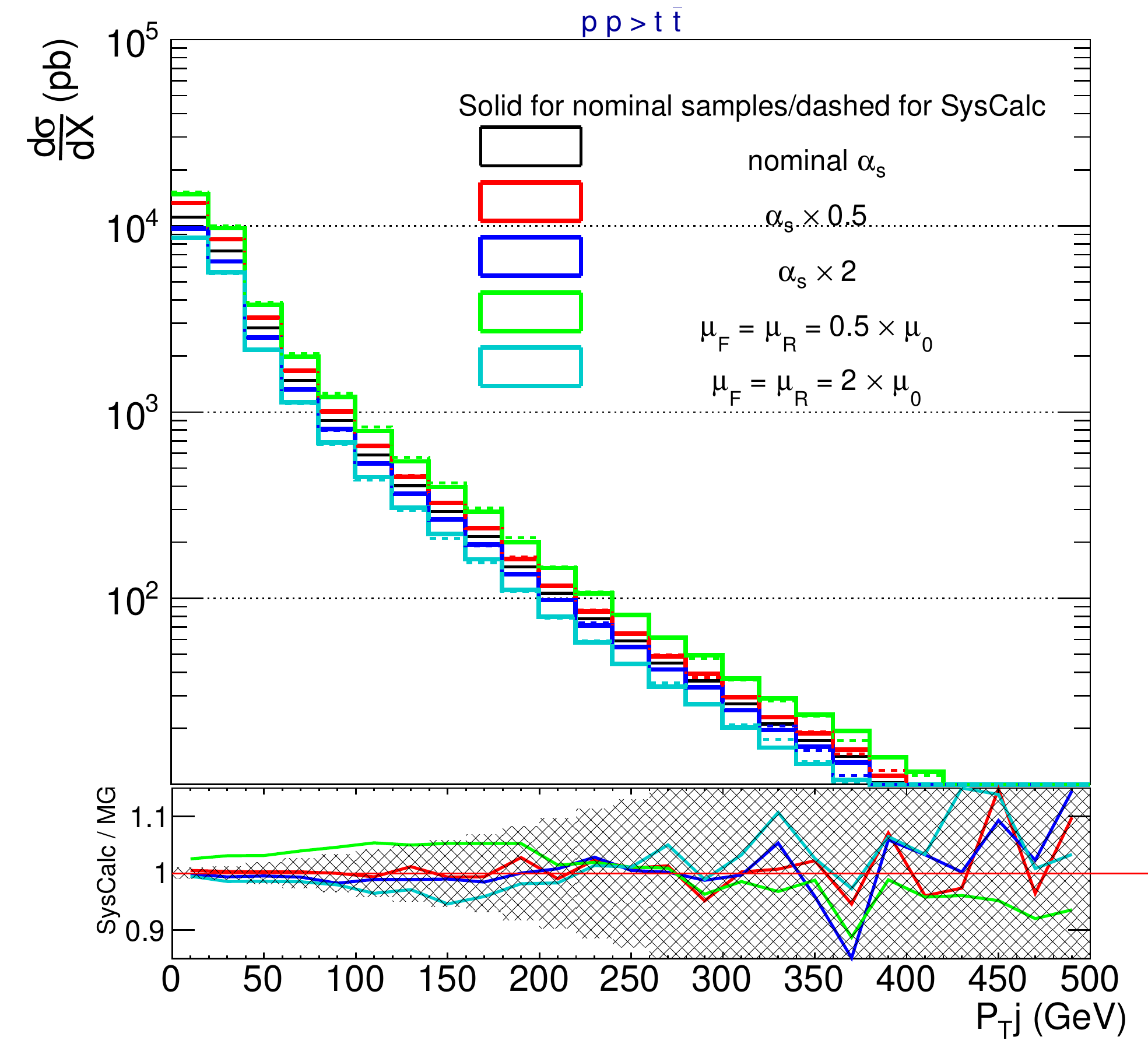}
          \includegraphics[width=0.5\linewidth, height = 0.3\textheight, keepaspectratio=true]{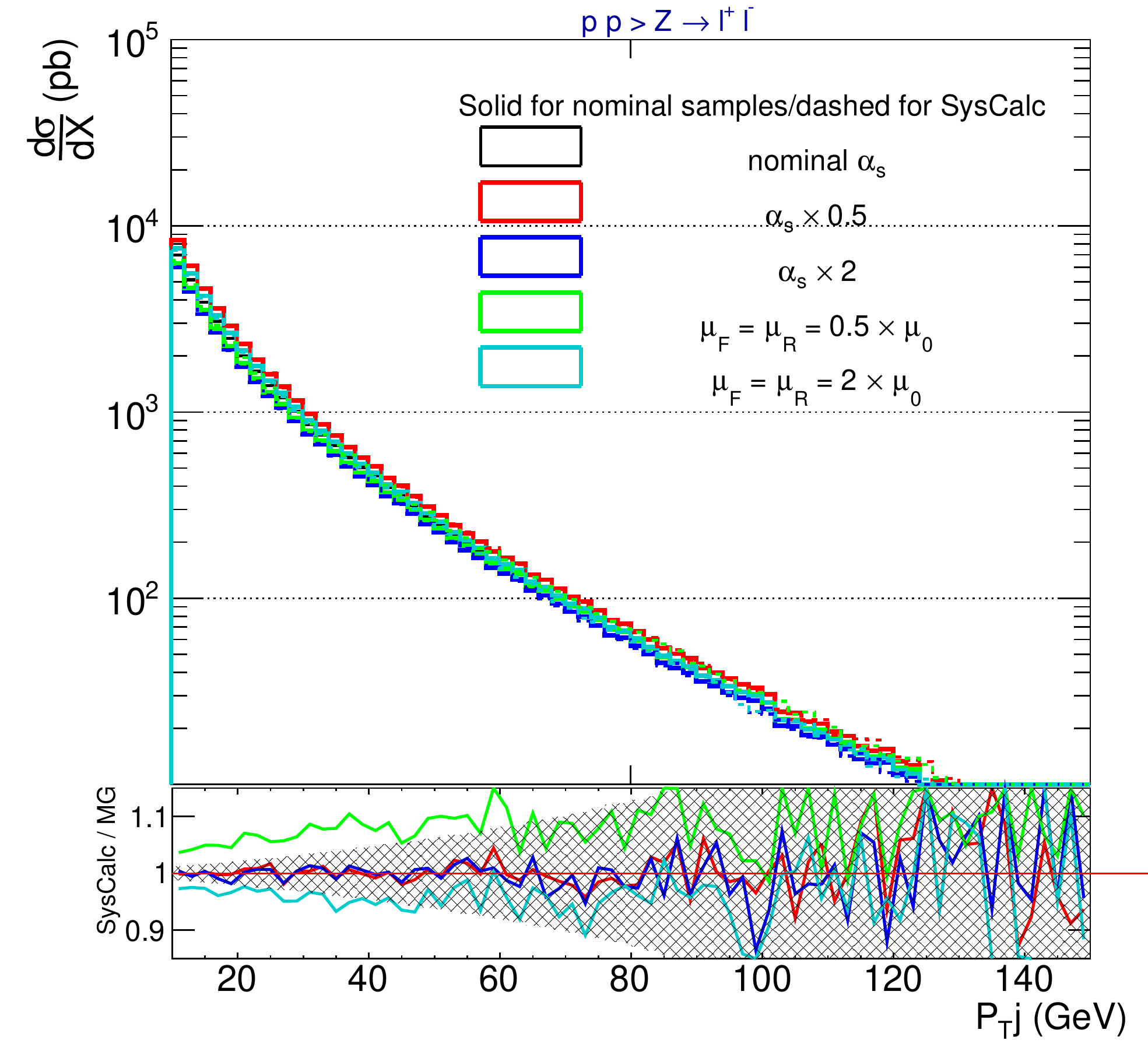}
    
          \includegraphics[width=0.5\linewidth, height = 0.3\textheight, keepaspectratio=true]{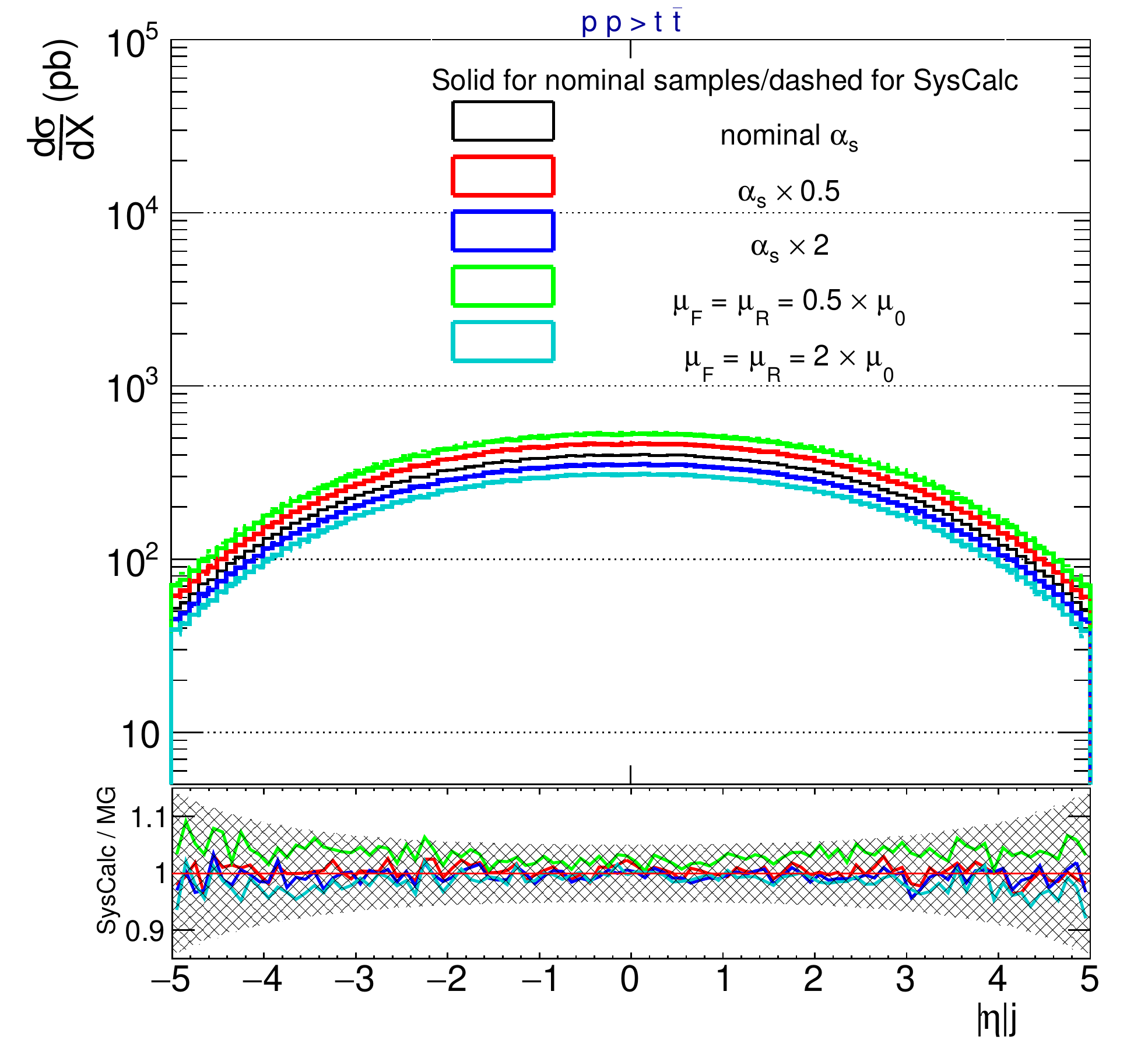}
          \includegraphics[width=0.5\linewidth, height = 0.3\textheight, keepaspectratio=true]{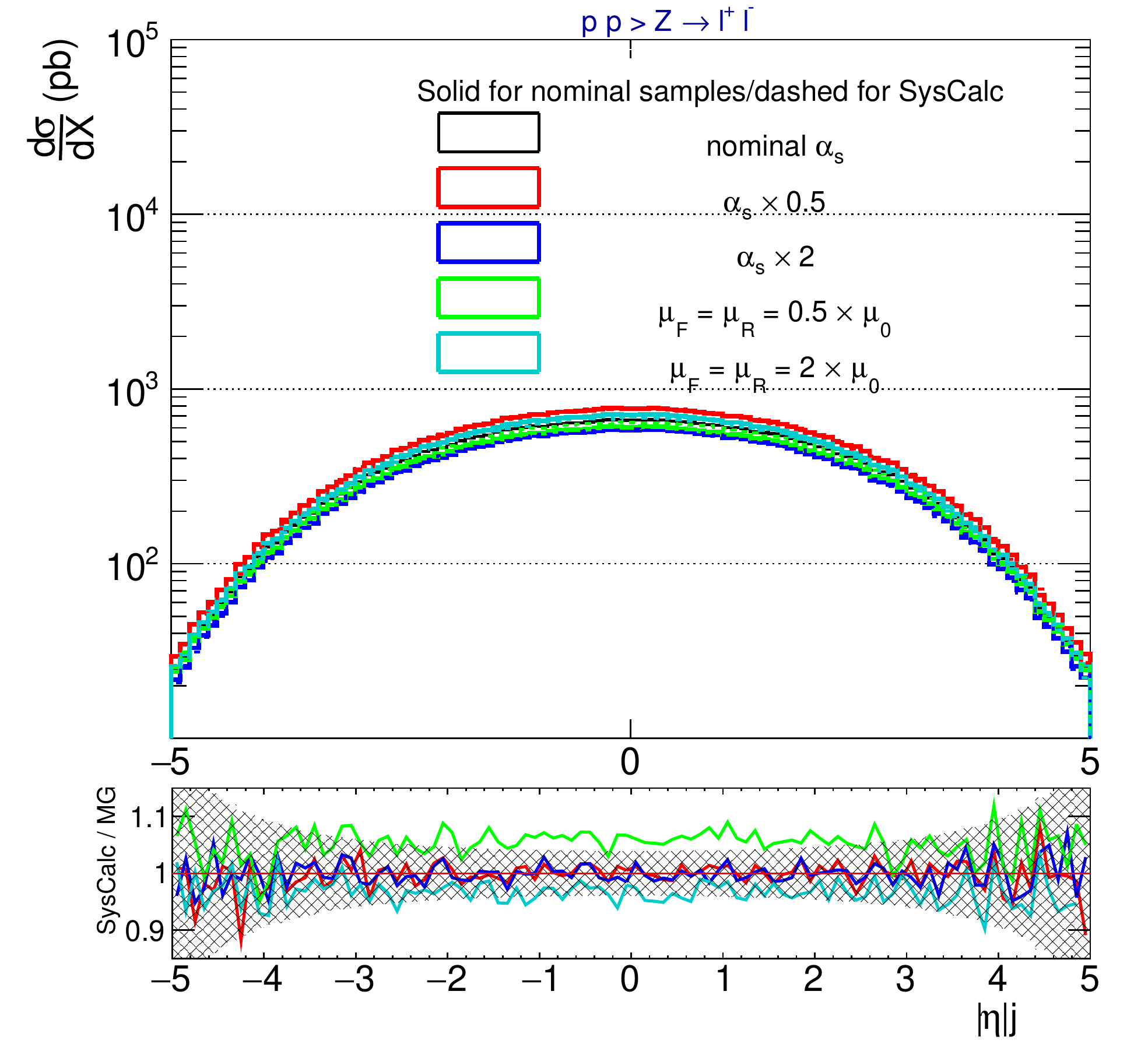}
      
             \end{minipage}}
             
 \caption{Distributions from $ p p  \to t \bar t ~+0,1j$ (left column) and $ p p \to Z ~+0,1j, ~Z \to l^+ l^-$ (right column) processes of the invariant mass of the $t(\bar t)\rm{~and~the~}Z$-boson respectively, as well as the \pt\ and the $|\eta|$ distributions of the ISR jet for various choices of the $\alpha_s$ the $\mu_{\rm F}$ and the $\mu_{\rm R}$ scales. The comparison is between the dedicated samples and the weighted events from \syscalc\ derived from the central sample (black solid line). The bottom on each distribution represents the ratio between the dedicated sample and the weighted events. The shaded area corresponds to the statistical uncertainty from the central sample. The small bias that appears in the middle and in the lower rows from variations of the $\mu_{\rm F}, \mu_{\rm R}$ scales is due to different starting scale of the parton shower. The details are discussed in  the text.}
  \label{fig:scales_matched} 

 \end{figure}

In Fig.~\ref{fig:scales_matched}  the invariant mass of the top quark pair and the Z-bosons, as well as the \pt\ and the $|\eta|$ distributions of the ISR jet for the different scale variations  from the dedicated samples are compared against the weighted events derived with \syscalc\ from the central sample. The agreement  is found to be within the statistical uncertainties, with one exception discussed below. Similarly the plots for the scales variations are presented in Fig.~\ref{fig:pdf_matched}. \\

One can notice that there is a small bias of the order of few \% level on the $\mu_{\rm F}$ and the $\mu_{\rm R}$ variations which is more prominent in the ISR kinematic related properties. 
This can be attributed to the fact that \pythia\ uses as the starting scale for the parton shower the scale that is already written in the LHE event record (parameter \texttt{SCALUP} in the \texttt{LHEv3} format) 
which  determines the maximum hardness of the first shower emission. 
For different choices of the $\mu_{\rm{F}}$ and the $\mu_{\rm{R}}$ scales, the \texttt{SCALUP} will change accordingly, 
but since \pythia\ runs only once on the nominal sample which holds the weights from \syscalc, thus it is not possible to have information 
on events with various \texttt{SCALUP} variations. Since this affects the hardness of the parton shower radiation, it is then expected to cause a difference in the final matched sample.\\

In order to check this effect, the following steps were carried out :
\begin{itemize}
\item The nominal sample (i.e. which has $\mu_{\rm{F}} = \mu_{\rm{R}} = 1$) was processed
after modifying the relevant parameters that control the \texttt{SCALUP} parameter\footnote{this is achieved by defining appropriately the \texttt{TimeShower:pTmaxMatch, TimeShower:pTmaxFudge, SpaceShower:pTmaxMatch, SpaceShower:pTmaxFudge} parameters in the \pythia\ card.}, so that  two new samples were derived; both of them having the nominal $\mu_{\rm{F}},\mu_{\rm{R}}$ scales set in,
 but with different $\texttt{SCALUP}$ corresponding to the $\times 2$ or $\times 0.5$ variations.   
 
 \item The dedicated samples generated with the variation on $\mu_{\rm{F}} = \mu_{\rm{R}} = 0.5$ and 2, were processed appropriately so that their $\texttt{SCALUP}$ corresponded to
 that of the nominal ($\mu_{\rm{F}} = \mu_{\rm{R}} =1$) samples.  
 \item As the above is equivalent to having separate samples corresponding to $\mu_{\rm{F}} = \mu_{\rm{R}} = 0.5~(2)$ 
 without modifying their \texttt{SCALUP}, dedicated samples with these settings were produced as well, with the matching efficiency found to be $\sim 27 (35)\%$ for the $\times 2~ (\times 0.5)$ variations samples correspondingly.

  \end{itemize}
 All of the above samples were compared against the prediction obtained from \syscalc.

\begin{figure}[!htp]
 \captionsetup[subfigure]{labelformat=empty}
    \centering
    \subfloat[]{
       \begin{minipage}{\linewidth}
        \includegraphics[width=0.5\linewidth, height = 0.5\textheight, keepaspectratio=true]{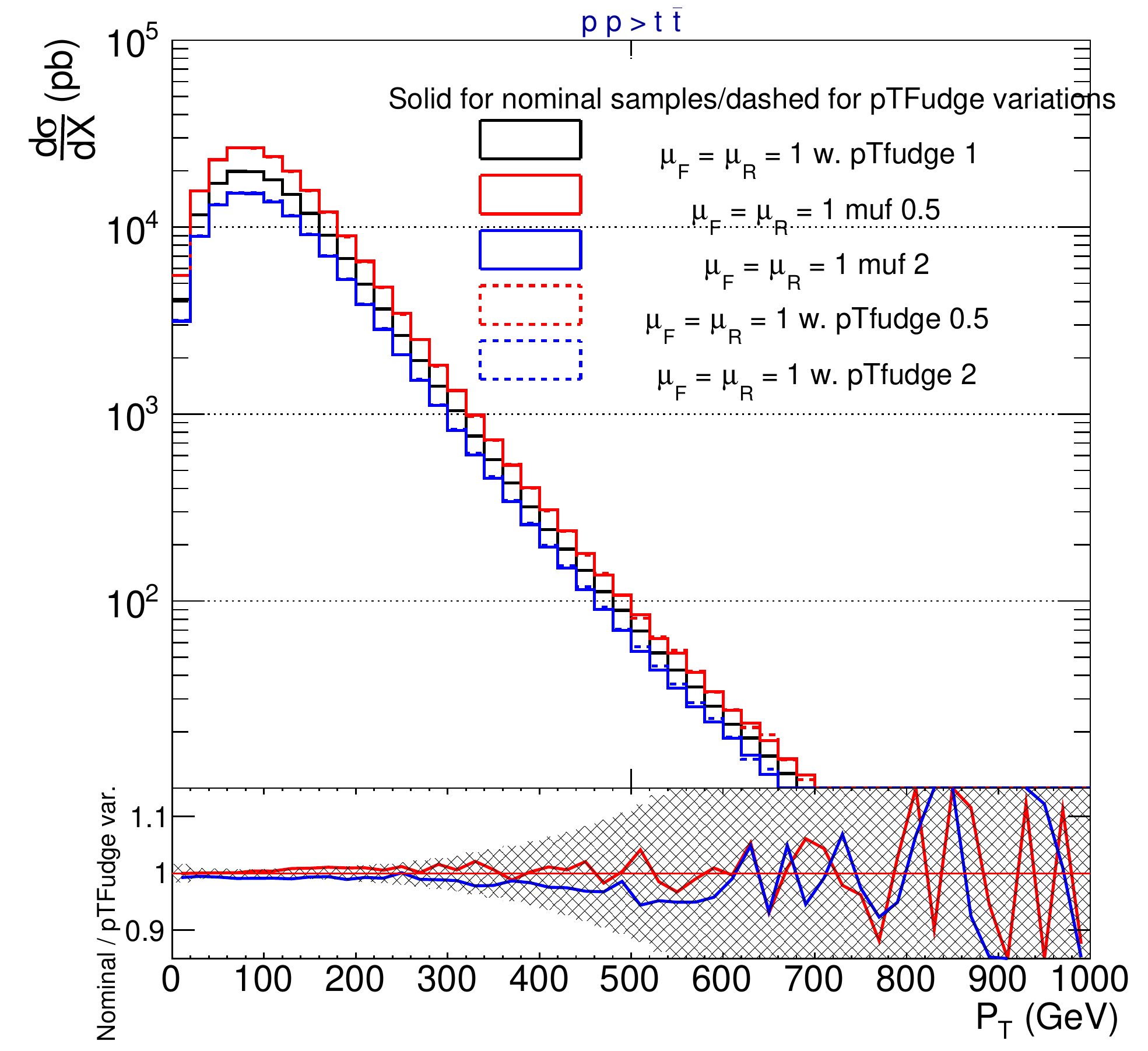}
            \includegraphics[width=0.5\linewidth, height = 0.5\textheight, keepaspectratio=true]{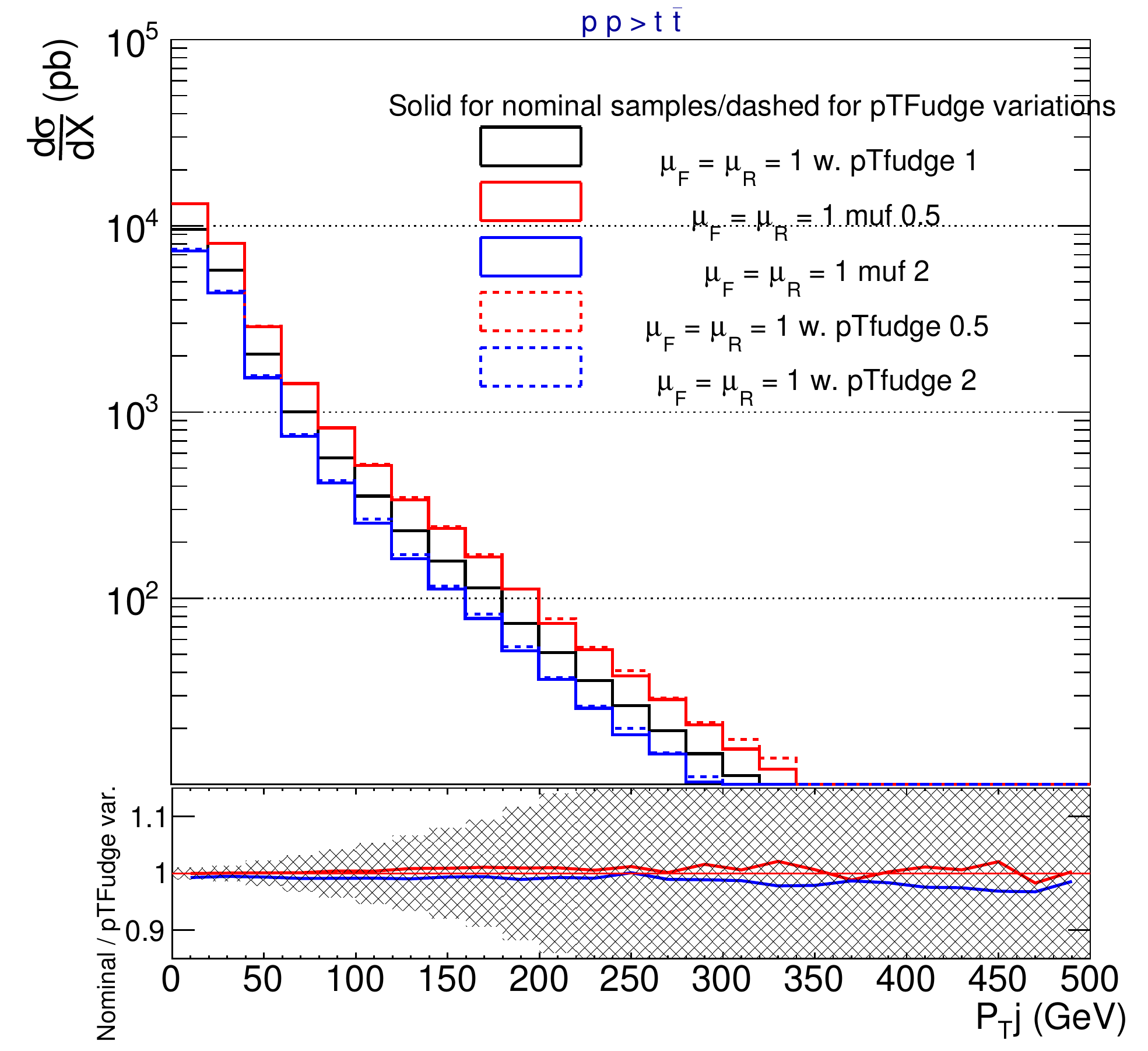}

             \end{minipage}}
             
 \caption{The \pt\ distribution from the $ p p  \to t \bar t ~+0,1j$ process of the $t(\bar t)$ (left) and from the ISR jet (right) and for different choices of the \texttt{SpaceShower:pTmaxFudge} option in \pythia.}
  \label{fig:fudge} 

 \end{figure}

 Further, for this exercise in order to assess the effects related to ISR and FSR radiation only, we veto all radiation related to the $t\bar{t}$ system and we consider jets at generator level only. 
 The \pt\ of these jets  are shown in Fig.~\ref{fig:fudge}; there is good agreement between the 
 dedicated and the systematic variation samples. 
 In summary, the choice of the \texttt{SCALUP} has a visible effect on the kinematics of the particles which is amplified for the ISR related quantities, 
 which can be well covered at analysis level by an extra systematic of the order of $5-10\%$ (for those quantities).  
\\

 \subsection{Future developments}

\syscalc\ can include the weight associated to different merging scales in the MLM matching/merging mechanism (from output provided by the \textsc{Pythia 6} package of \textsc{pythia-pgs package}). In that case, the parton shower keeps track of the scale of the first emission and applies then a veto to account for the minimal allowed value for the matching scale according to the cut performed at parton-level. \syscalc\ will then test for each of the values specified in the parameter \emph{matchscale} if the event passes the MLM criteria or not. If it does not,  a zero weight is associated to the events, while if it does,  a weight of 1 is kept. As a reminder,  these weights are the equivalent of having a (approximate) Sudakov form-factor and removing at the same time the double counting between the events belonging to different multiplicities. However, as at the time of writing this paper, \textsc{Pythia6} is no longer supported and the described functionality is not yet supported in \pythia,  this functionality cannot  be fully tested currently at the moment.\\

\section{Practical guide to use \syscalc}
\label{guide}
The  requirements of the \syscalc\ package as inputs are:

\begin{itemize}
\item A systematics file (which can be generated by MadGraph 5 v.1.6.0 or later) where the generated events will be further processed so to calculate the weights. 
\item The presence of \texttt{LHAPDF}~\cite{LHAPDF} installed in the working environment\footnote{Instructions on how to install LHAPDF can be found in the following link: \texttt{http://lhapdf.hepforge.org/lhapdf6/install.html}}. Once \texttt{LHAPDF} is installed, it is very important to properly set the environment variables (assuming that \texttt{LHAPDF} was installed in the \texttt{/local} directory
\\
\texttt{
export PATH=\$PWD/local/bin:\$PATH\\
export LD\_LIBRARY\_PATH=\$PWD/local/lib:\$LD\_LIBRARY\_PATH\\
export PYTHONPATH=\$PWD/local/lib64/python2.6/site-packages:\$PYTHONPATH
}

\item A configuration file (i.e. simple text file) specifying the parameters to be varied as described later.
\end{itemize}

\subsection{Installation}
In any system with \texttt{bazaar} available:
\texttt{\\
     bzr branch lp:~mgtools/mg5amcnlo/SysCalc \\
     cd SysCalc \\
     make\\
     }

 \subsection{Configuration file}
Before running \syscalc\ a configuration card with the parameters to be varied is needed. Below follows such an example on how to vary the central scale, the $\alpha_s$ and the PDF set:

\texttt{\\
\# Central scale factors \\
scalefact: \\
0.5 1 2\\
\# Scale correlation \\
\# Special value -1: all combination (N**2)\\
\# Special value -2: only correlated variation\\
\# Otherwise list of index N*fac\_index + ren\_index \#index starts at 0\\
scalecorrelation: \\
-1 \\
\#Emission scale factors\\
alpsfact:
0.5 1 2 \\
\#PDF sets and number of members\\
PDF:\\
CT10nlo\\
}
\\
If your are interested to simply run on one member of the PDF grid, simply replace the last line with this: \\ 
\texttt{\\
CT10nlo 1
} 
\\

Numbering of the PDF members starts from 1, so the above line will result to consider only the first member (ie the central one) of a given PDF grid.

Also, a LHE v3 file is needed which has to be generated with \mg\ with the option:
\\
\newline
\texttt{T = sys\_calc}. \\

 \subsection{Executing}

The one-line syntax to execute \syscalc\ is: 
\\

\texttt{./syscalc input.lhe syscalc\_parameters.dat out.lhe}
\\

The output code  follows the LHEF v3 format. The following block appears in the header of the output file. Assuming the above configuration card, the output will look like:

\texttt{
\\
<header>\\
<initrwgt>\\
  <weightgroup type="Central scale variation" combine="envelope">\\
    <weight id="1" MUR="0.5" MUF="0.5" PDF="10042"> mur=0.5 muf=0.5 </weight>\\
    <weight id="2" MUR="0.5" MUF="1" PDF="10042"> mur=0.5 muf=1 </weight>\\
    <weight id="3" MUR="0.5" MUF="2" PDF="10042"> mur=0.5 muf=2 </weight>\\
    <weight id="4" MUR="1" MUF="0.5" PDF="10042"> mur=1 muf=0.5 </weight>\\
    <weight id="5" MUR="1" MUF="1" PDF="10042"> mur=1 muf=1 </weight>\\
    <weight id="6" MUR="1" MUF="2" PDF="10042"> mur=1 muf=2 </weight>\\
    <weight id="7" MUR="2" MUF="0.5" PDF="10042"> mur=2 muf=0.5 </weight>\\
    <weight id="8" MUR="2" MUF="1" PDF="10042"> mur=2 muf=1 </weight>\\
    <weight id="9" MUR="2" MUF="2" PDF="10042"> mur=2 muf=2 </weight>\\
  </weightgroup>
  <weightgroup name="Emission scale variation" combine="envelope">
    <weight id="10" ALPSFACT="0.5" MUR="1" MUF="1" PDF="10042"> alpsfact=0.5</weight>
    <weight id="11" ALPSFACT="1" MUR="1" MUF="1" PDF="10042"> alpsfact=1</weight>
    <weight id="12" ALPSFACT="2" MUR="1" MUF="1" PDF="10042"> alpsfact=2</weight>
  </weightgroup>
      <weightgroup name="CT10nlo" combine="hessian">
    <weight id="13" MUR="1" MUF="1" PDF="11000"> Member 0 of sets CT10nlo</weight>\\
    <weight id="14" MUR="1" MUF="1" PDF="11001"> Member 1 of sets CT10nlo</weight>\\
    <weight id="15" MUR="1" MUF="1" PDF="11002"> Member 2 of sets CT10nlo</weight>\\
    <weight id="16" MUR="1" MUF="1" PDF="11003"> Member 3 of sets CT10nlo</weight>\\
    <weight id="17" MUR="1" MUF="1" PDF="11004"> Member 4 of sets CT10nlo</weight>\\
    <weight id="18" MUR="1" MUF="1" PDF="11005"> Member 5 of sets CT10nlo</weight>\\
    <weight id="19" MUR="1" MUF="1" PDF="11006"> Member 6 of sets CT10nlo</weight> \\
    ....\\
     <weight id="64" MUR="1" MUF="1" PDF="11051"> Member 51 of sets CT10nlo</weight>\\
    <weight id="65" MUR="1" MUF="1" PDF="11052"> Member 52 of sets CT10nlo</weight>\\
  </weightgroup>
</initrwgt>\\
</header>\\
}

While for each event, the following information will be written: \\ 
\texttt{
\\
</mgrwt>
<rwgt>
  <wgt id="1">0.128476</wgt>\\
  <wgt id="2">0.119447</wgt>\\
  <wgt id="3">0.111059</wgt>\\
  <wgt id="4">0.0946238</wgt>\\
  <wgt id="5">0.087974</wgt>\\
  <wgt id="6">0.0817961</wgt>\\
  <wgt id="7">0.0716918</wgt>\\
  <wgt id="8">0.0666536</wgt>\\
  <wgt id="9">0.0619729</wgt>\\
  <wgt id="10">0.087974</wgt>\\
  <wgt id="11">0.087974</wgt>\\
  <wgt id="12">0.087974</wgt> \\
   ...\\
  <wgt id="64">34893.5</wgt>\\
  <wgt id="65">41277</wgt>\\
</rwgt>
 }

\section{Conclusion}
\label{close}
In this paper, we presented the \syscalc\ package,  a C++ written tool capable of calculating weights for certain theoretical systematic uncertainties. 
The very existence of this tool tackles the problem that many physics analyses face when they have to deal with calculation of different yet important
theoretical systematic uncertainties but  struggle with the generation of large simulated dedicated samples. 
With the use of \syscalc, one is able to have  a systematics sample with similar statistical power as for the nominal one as the main idea is based on derivation of dedicated weights, one for 
each source of the considered theoretical systematics. 
Further, the code is very fast and thus ideal for a large scale MC production, while the final weights are written in dedicated tags in $\texttt{ASCII}$ using the $\texttt{LHEFv3}$ format. 
We presented a first level of validation of \syscalc's main characteristics along with a practical guide on how to use it.
\\

Near future developments include a full integration with \pythia\ in order to provide the matching scales variations on top of the currently available ones.

 \section{Acknowledgements} 
 The authors would like to warmly thank Olivier Mattelaer and Steven Mrenna for their continuous support, cooperation and fruitful discussions.

 %   $\to$ Appendix format to store the information

%\section{Internal LHE format for SysCalc}

%\bibliography{mybibfile,valentin}

\end{document}